%% file: main.tex
\documentclass[conf]{new-aiaa} %for conference papers
\usepackage[utf8]{inputenc}

\usepackage{graphicx}

\usepackage{amsmath}
\usepackage{dsfont}
\usepackage{hhline}
\usepackage{booktabs}
\usepackage{tabularx}
\usepackage{multirow}
\usepackage{boldline}
\usepackage{bm}
\usepackage{makecell}
\usepackage{indentfirst}
\usepackage{diagbox}
\newcommand\nhline[1]{\noalign{\hrule height #1}}
\input{PackagesCommands}
\usepackage[colorinlistoftodos,textwidth=1 in,textsize=footnotesize]{todonotes}
\title{Model-free Adaptive Output Feedback Vibration Suppression \\  in a Cantilever Beam}

\author{
Juan Augusto Paredes Salazar\footnote{Postdoctoral Research Fellow, Department of Mechanical Engineering, University of Maryland, Baltimore County, 1000 Hilltop Circle, Baltimore, MD 21250. \texttt{japarede@umbc.edu}}, 
Ankit Goel\footnote{Assistant Professor, Department of Mechanical Engineering, University of Maryland, Baltimore County, 1000 Hilltop Circle, Baltimore, MD 21250. \texttt{ankgoel@umbc.edu}}
}

\begin{document}
\maketitle

\begin{abstract}
    This paper presents a model-free adaptive control approach to suppress vibrations in a cantilevered beam excited by an unknown disturbance.
    % with a displacement or acceleration measurement at an arbitrary point. 
    % Motivated by the flutter problem, 
    % A lumped parameter model of the system is considered to model the cantilevered beam under excitation.
    The cantilevered beam under harmonic excitation is modeled using a lumped parameter approach. 
    Based on retrospective cost optimization, a sampled-data adaptive controller is developed to suppress vibrations caused by external disturbances. 
    Both displacement and acceleration measurements are considered for feedback.
    %
    %Since the acceleration sensor cannot provide a displacement measurement, a novel filter is developed to extract key displacement information from the acceleration data.
    %
    Since acceleration measurements are more sensitive to spillover, which excites higher frequency modes, a filter is developed to extract key displacement information from the acceleration data and enhance suppression performance.
    The vibration suppression performance is compared using both displacement and acceleration measurements. 
\end{abstract}

\section{Introduction}\label{sec:intro}

Vibrations caused by external disturbances occur in several engineering applications, resulting in decreased performance and potentially leading to instability.
Active vibration control techniques have been proposed to address this issue \cite{alkhatib2003, vasques2006, caresta2011, aridogan2015, preumont2018, cunha2023, li2023}.
In particular, control vibration suppression has been applied to 
marine vehicles and structures for vibrations induced by machinery \cite{daley2004, zhang2017, soni2020},
building structures for vibrations induced by strong winds and earthquakes \cite{shiba1998, li2020, ramirez2022},
and aerospace vehicles for vibrations in satellites and flexible wings \cite{hu2005, hu2009, he2015, tsushima2018, prakash2016, bloemers2024}.
Numerous techniques have been implemented for this purpose, including 
LQR \cite{zhang2008, schulz2013, zhang2024}, 
robust control \cite{souza2019,fan2019,wang2022},
and adaptive control \cite{qiu2006, landau2020, batista2021, saeed2022, wang2023, liu2024}.

In this manuscript, we propose the implementation of Retrospective Cost Adaptive Control (RCAC) \cite{rahmanCSM2017}, a model-free output feedback technique, for vibration suppression.
RCAC has been previously implemented to suppress oscillations in vibrational and self-excited systems \cite{mohseni2022, paredes2022, paredes2023}.
A cantilever beam system under external disturbance is considered a testbed for its ubiquitous presence in the active vibration control literature.

\begin{figure}[h]
    \centering
    \includegraphics[width = 0.5\columnwidth]{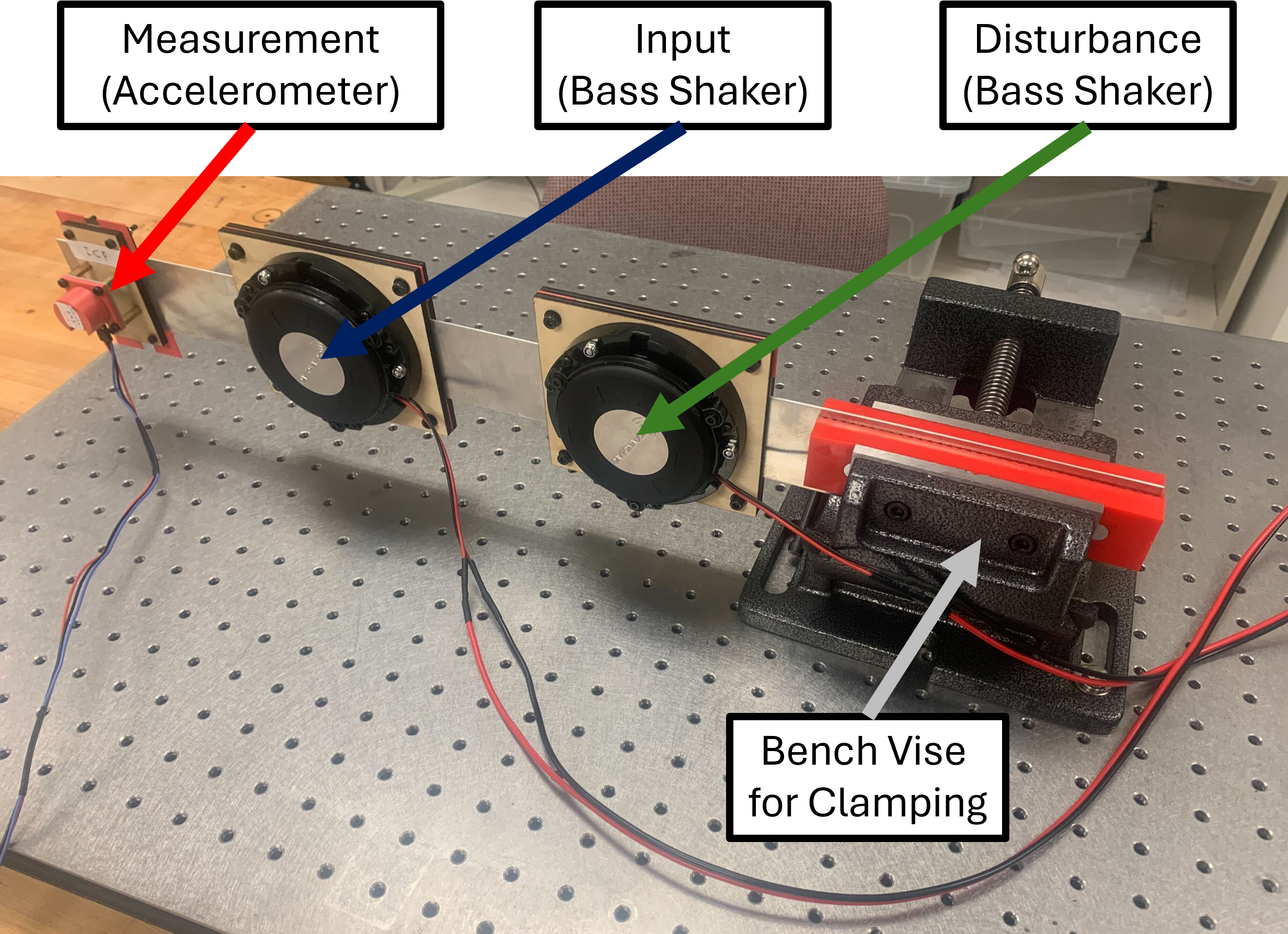}
    \caption{Cantilever beam laboratory setup.}
    \label{fig:cantilever_exp}
\end{figure}

The structural model considered in this paper is motivated by the cantilever beam shown in Figure \ref{fig:cantilever_exp}.
A bass shaker is used to harmonically excite the cantilever and the objective is to minimize the displacement at the sensor location. 
The input is applied by another bass shaker mounted at a different point on the cantilever.
Note that the measurement, actuation, and disturbance are not colocated in this example. 
Both position and acceleration measurements are considered for feedback. 

Since acceleration measurements are more sensitive to spillover, which results in higher frequency modes being excited by the controller \cite{balas1982,hyland1993,hong1998,alkhatib2003,dong2014}, a filter is developed to extract key displacement information from the acceleration data and improve suppression performance \cite{zhu2006, an2015, mounesisohi2017, wei2019, huang2021}.
The vibration suppression performance of RCAC is compared using both displacement and acceleration measurements. 

The contents of the paper are as follows.
Section \ref{sec:LPM} introduces a lumped parameter model for the cantilever beam system, motivated by the cantilever beam shown in Figure \ref{fig:cantilever_exp}.
Section \ref{sec:control} provides an overview of the control system, including a review of RCAC, the sampled-data implementation of RCAC that interfaces the discrete-time controller with the continuous-time system, and the signal conditioning filters used for displacement and acceleration measurements.
Section \ref{sec:simulations} provides the results of numerical simulations in which RCAC is implemented to suppress vibrations in the lumped parameter model developed in Section \ref{sec:LPM}.
Section \ref{sec:conclusions} provides conclusions.

{\bf Notation:}
$\BBR \isdef (-\infty,$ $\infty),$ 
$\BBR^{\ge0} \subset \BBR \isdef [0,$ $\infty),$ 
$\BBZ \isdef \{\ldots, -2, -1, 0, 1, 2, \ldots\},$
and $\Vert\cdot\Vert$ denotes the Euclidean norm on $\BBR^n.$
$x_{(i)}$ denotes the $i$th component of $x\in\BBR^n.$
$\bfq \in \BBC$ is the forward-shift operator.
For all $i \in \{1, \ldots, n\},$ let $e_{n,i} \isdef \matl e_{n,i (1)} & e_{n,i (2)} & \cdots & e_{n,i (n)} \matr^\rmT \in \BBR^n,$ such that
\begin{equation*}
    e_{n,i (j)} \isdef    \begin{cases}
                            1, & \mbox{if } i = j,\\
                            0, & \mbox{otherwise}.
                        \end{cases}
\end{equation*}
The symmetric matrix $P\in\BBR^{n\times n}$ is positive semidefinite (resp., positive definite) if all of its eigenvalues are nonnegative (resp., positive).
$\vek X\in\BBR^{nm}$ denotes the vector formed by stacking the columns of $X\in\BBR^{n\times m}$, and 
$\otimes$ denotes the Kronecker product.
$I_n$ is the $n \times n$ identity matrix, $0_{n\times m}$ is the $n\times m$ zeros matrix, and $\mathds{1}_{n\times m}$ is the $n\times m$ ones matrix.

\section{Lumped Parameter Model}\label{sec:LPM}

\begin{figure}[h]
    \centering
    \includegraphics[width = 0.7\columnwidth]{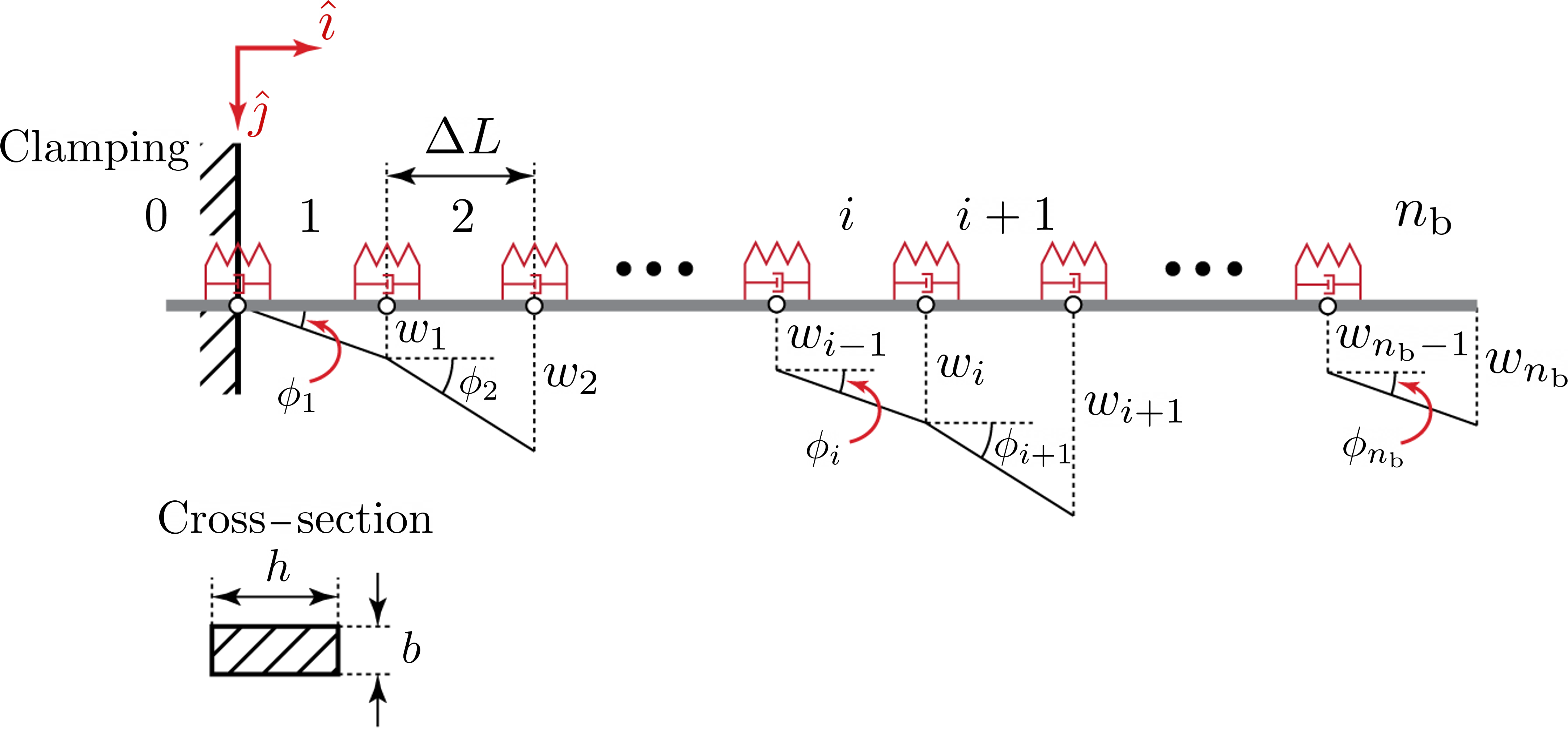}
    \caption{Lumped parameter model of a cantilever beam.}
    \label{fig:LP_cantilever}
\end{figure}

In this section, a lumped parameter (LP) model of a cantilever beam is derived based on the methodology shown in \cite{tuuma2012}.
A LP model of a cantilever beam clamped to a wall of length $L,$ cross-section width $b,$ cross-section height $h,$ and mass $m$ is obtained by dividing the beam into $n_\rmb$ elements of length $\Delta L \isdef L / n_\rmb,$ as shown in Figure \ref{fig:LP_cantilever}.
Let axes $\hat{\imath}$ and $\hat{\jmath}$ be perpendicular and parallel to the wall, respectively, as shown in Figure \ref{fig:LP_cantilever}, and note that the beam is aligned with $\hat{\imath}$ when it is unforced.
Let $p_0$ be the clamping point on the wall.
For all $i \in \{1, \ldots, n_\rmb\},$ let $p_i$ be the the farthest point of the $i-$th element from $p_0,$ let $w_i$ be the displacement of $p_i$ in the direction of $\hat{\jmath}$ , let $w_{\rmc, i}$ be the displacement of the center of mass the $i-$th element in the direction of $\hat{\jmath},$ such that
\begin{equation}
    w_{\rmc, i} \isdef \begin{cases}
                            \frac{w_1}{2}, & \mbox{if } i = 1,\\
                            \frac{w_{i-1} + w_i}{2}, & \mbox{if } i \in \{2, \ldots, n_\rmb\},
                        \end{cases} \label{eq:w_c}
\end{equation}
let $\phi_i$ be the angle of rotation of the $i-$th element relative to $\hat{\imath},$ and let $f_i$ be the external force applied to $p_i$ in the direction of $\hat{\jmath}.$ 
Assume that $\phi_1, \ldots, \phi_{n_\rmb} $ are small.
Then,
\begin{equation}
    \phi_i = \begin{cases}
                \phi_1 = \frac{w_1}{\Delta L} & \mbox{if } i = 1,\\
                \phi_i = \frac{w_i - w_{i-1}}{\Delta L}  & \mbox{if } i \in \{2, \ldots, n_\rmb\}.
             \end{cases} \label{eq:phi}
\end{equation}
For all $j \in \{0, \ldots, n_\rmb-1\},$ let the angle between adjacent elements at point $p_j$ be given by
\begin{equation}
    \Delta \phi_j \isdef \begin{cases}
                      \phi_1 = \frac{w_1}{\Delta L} & \mbox{if } j = 0,\\
                      \phi_2 - \phi_1 = \frac{w_2 - 2 w_1}{\Delta L} & \mbox{if } j = 1,\\
                      \phi_{j+1} - \phi_j = \frac{w_{j+1} - 2 w_j + w_{j-1}}{\Delta L} & \mbox{if } j \in \{2, \ldots, n_\rmb-1\}.
                      \end{cases} \label{eq:Delta_phi}
\end{equation}

Next, assuming that the force of gravity has no effect on the displacement of any element of the beam, the total potential energy $U$ and the total kinetic energy $T$ relative to $p_0$ are given by
\begin{align}
    U &= \sum_{j = 0}^{n_\rmb - 1} \frac{1}{2} K_\phi \left( \Delta \phi_j \right)^2, \label{eq:potential_energy} \\
    T &= \sum_{i = 1}^{n_\rmb} \frac{1}{2} \Delta m \dot{w}_{\rmc, i}^2 + \frac{1}{2} J \dot\phi_i^2, \label{eq:kinetic_energy}
\end{align}
where $\Delta m \isdef m / n_\rmb,$ 
%
%\begin{equation*}
    $K_\phi \isdef \frac{E h b^3 }{4 \Delta L},$
%\end{equation*}
%
is the stiffness between each element, $E$ is Young's modulus corresponding to the beam material,
and $J \isdef \Delta m\frac{\Delta L^2 + b^2}{12}$ is the inertia of each element along an axis perpendicular to the plane.
Then, it follows from \eqref{eq:w_c} -- \eqref{eq:kinetic_energy} that, for all $i \in \{1, \ldots, n_\rmb\},$
\begin{align}
    \frac{\partial U}{\partial w_i} &= \begin{cases} 
                                            \frac{K_\phi}{\Delta L ^2} \left( 6 w_1 - 4 w_2 + w_3 \right) & \mbox{if } i = 1,\\[1.5ex]
                                            \frac{K_\phi}{\Delta L ^2} \left( -4 w_1 + 6 w_2 - 4 w_3 + w_4 \right) & \mbox{if } i = 2,\\[1.5ex]
                                            \frac{K_\phi}{\Delta L ^2} \left( w_{i - 2} - 4 w_{i - 1} + 6 w_i - 4 w_{i+1} + w_{i + 2} \right) & \mbox{if } i \in \{3, \ldots, n_\rmb -2\},\\[1.5ex]
                                            \frac{K_\phi}{\Delta L ^2} \left( w_{n_\rmb - 3} - 4 w_{n_\rmb - 2} + 5 w_{n_\rmb - 1} - 2 w_{n_\rmb} \right) & \mbox{if } i = n_\rmb - 1,\\[1.5ex]
                                            \frac{K_\phi}{\Delta L ^2} \left( w_{n_\rmb - 2} - 2 w_{n_\rmb - 1} + w_{n_\rmb} \right) & \mbox{if } i = n_\rmb,
                                        \end{cases}, \label{eq:dw_Potential} \\
    \frac{\partial}{\partial t} \left( \frac{\partial T}{\partial \dot{w}_i} \right) &=
    \begin{cases}
        \gamma_1 \ddot{w}_1 + \gamma_2 \ddot{w}_2 & \mbox{if } i = 1, \\[1ex]
        \gamma_2 \ddot{w}_{i - 1} + \gamma_1 \ddot{w}_i + \gamma_2 \ddot{w}_{i + 1} & \mbox{if } i \in \{2, \ldots, n_\rmb - 1 \}, \\[1ex]
        \gamma_2 \ddot{w}_{n_\rmb - 1} + \frac{\gamma_1}{2} \ddot{w}_{n_\rmb} & \mbox{if } i = n_\rmb,
    \end{cases}  \label{eq:dt_dwdot_Kinetic} 
\end{align}
where
\begin{align*}
    \gamma_1 &= \frac{\Delta m}{2} \left( 1 + \frac{1}{3} \left( 1 + \left( \frac{b}{\Delta L} \right)^2 \right) \right), \
    \quad
    \gamma_2 = \frac{\Delta m}{4} \left( 1 - \frac{1}{3} \left( 1 + \left( \frac{b}{\Delta L} \right)^2 \right) \right).
\end{align*}
Hence, it follows from \eqref{eq:dw_Potential},\eqref{eq:dt_dwdot_Kinetic}, and the Euler-Lagrange equations that, for all $i \in \{1, \ldots, n_\rmb\},$ the equations of motion of the cantilever system are given by
\begin{equation}
    \frac{\partial}{\partial t} \left( \frac{\partial T}{\partial \dot{w}_i} \right) + \frac{\partial U}{\partial w_i} = f_i.\label{eq:eom1} 
\end{equation}
Rewriting \eqref{eq:eom1} into a matrix differential equation yields
\begin{equation}
    M \ddot{w} + K w = f,\label{eq:eom2} 
\end{equation}
where
\begin{equation*}
   w \isdef \matl w_1 & \cdots & w_{n_\rmb} \matr^\rmT, \quad 
   f \isdef \matl f_1 & \cdots & f_{n_\rmb} \matr^\rmT,
\end{equation*}
\begin{equation*}
   M = \matl \gamma_1 & \gamma_2 & 0 & \cdots & \cdots & \cdots & \cdots \\ \vdots & \vdots & \vdots & \vdots & \vdots & \vdots & \vdots \\ \cdots & 0 & \gamma_2 & \gamma_1 & \gamma_2 & 0 & \cdots  \\ \vdots & \vdots & \vdots & \vdots & \vdots & \vdots & \vdots \\ \cdots & \cdots & \cdots & \cdots & 0 & \gamma_2 & \frac{\gamma_1}{2} \matr, \quad 
   K \isdef \frac{K_\phi}{\Delta L^2} \matl 6  & -4 & 1 & 0 & \cdots & \cdots & \cdots & \cdots & \cdots \\ -4 & 6 & -4 & 1 & 0 & \cdots & \cdots & \cdots & \cdots \\ \vdots & \vdots & \vdots & \vdots & \vdots & \vdots & \vdots & \vdots & \vdots \\ \cdots & 0 & 1 & -4 & 6 & -4 & 1 & 0 & \cdots \\ \vdots & \vdots & \vdots & \vdots & \vdots & \vdots & \vdots & \vdots & \vdots \\ \cdots & \cdots & \cdots & \cdots & 0 & 1 & -4 & 5 & -2 \\ \cdots & \cdots & \cdots & \cdots & \cdots & 0 & 1 & -2 & 1 \matr. 
\end{equation*}
To account for the presence of structural damping, \eqref{eq:eom2} is modified as
\begin{equation}
     M \ddot{w} + C_\rmR \dot{w} + K w = f,\label{eq:eom3} 
\end{equation}
where $C_\rmR \isdef \alpha M + \beta K$ is the Rayleigh damping matrix, and $\alpha, \beta \in \BBR$ determine the damping ratios corresponding to the eigenvalues of the cnatilever system matrix.

Next, let $i_\rmu, i_\rmd \in \{1, \ldots, n_\rmb\}$ be the beam elements at which the input and the disturbance are applied, respectively, let $u \in \BBR$ be an input force applied at $p_{i_\rmu},$ and let $d \in \BBR$ be an external disturbance applied at  $p_{i_\rmd},$
such that
\begin{equation}
    f = e_{n_\rmb,i_\rmu} u + e_{n_\rmb,i_\rmd} d. \label{eq:force} 
\end{equation}
Let $x \isdef \matl w^\rmT & \dot{w}^\rmT \matr^\rmT$
Then, it follows from \eqref{eq:eom3}, \eqref{eq:force} that the equations of motion of the cantilever system can be written in state space form as
\begin{equation}
    \dot{x} = A x + B_\rmu u + B_\rmd d, \label{eq:eom4} 
\end{equation}
where
\begin{equation*}
    A \isdef \matl 0_{n_\rmb \times n_\rmb} & I_{n_\rmb} \\ -M^{-1} K & -M^{-1} C_\rmR \matr, \quad B_\rmu \isdef \matl 0_{ n_\rmb \times 1} \\ M^{-1} e_{n_\rmb, i_\rmu} \matr , \quad B_\rmd \isdef \matl 0_{ n_\rmb \times 1} \\ M^{-1} e_{n_\rmb, i_\rmd} \matr.
\end{equation*}

Next, let $i_\rmy \in \{1, \ldots, n_\rmb\}$ be the beam element from which a measurement is obtained, let $y \in \BBR$ be a measurement obtained from $p_{i_\rmu}.$ 
In the case where a displacement measurement is available, $y = w_{i_\rmy},$ and
\begin{equation}
    y = C_{\rm disp} x,
\end{equation}
where
\begin{equation*}
    C_{\rm disp} \isdef \matl e_{n_\rmb, i_\rmy} & 0_{n_\rmb \times 1}^\rmT \matr^\rmT.
\end{equation*}
In the case where an acceleration measurement is available, $y = \ddot{w}_{i_\rmy},$ and
\begin{equation}
    y = C_{\rm acc} x + D_{{\rm acc},\rmu} u + D_{{\rm acc},\rmd} d,
\end{equation}
where
\begin{equation*}
    C_{\rm acc} \isdef \matl 0_{1 \times n_\rmb} & e_{n_\rmb, i_\rmy}^\rmT  \matr A, \quad D_{{\rm acc},\rmu} \isdef  \matl 0_{1 \times n_\rmb} & e_{n_\rmb, i_\rmy}^\rmT  \matr B_\rmu , \quad D_{{\rm acc},\rmd} \isdef  \matl 0_{1 \times n_\rmb} & e_{n_\rmb, i_\rmy}^\rmT  \matr B_\rmd.
\end{equation*}
The LP system considered in this manuscript is single-input, single-output (SISO).
The control objective for the LP system is to minimize the oscillations of $w_{i_\rmy},$ propagated by the disturbace signal $d$ by modulating the input signal $u,$ which is determined by a control algorithm based on the measurement signal, which can be a displacement measurement $w_{i_\rmy}$ or an acceleration measurement $\ddot{w}_{i_\rmy}.$

\section{Control}\label{sec:control}

In this section, the adaptive control algorithm and its implementation are introduced. 
Subsection \ref{subsec:rcac} provides a review of Retrospective Cost Adaptive Control (RCAC), a discrete-time, model-free, adaptice control algorithm. 
Subsection \ref{subsec:ctrl_impl} introduces the sampled-data implementation of RCAC to interface the discrete-time controller with the continuous-time system.
Subsection \ref{subsec:filter} provides an overview of the signal conditioning filters used to improve the performance of the adaptive controller.

\subsection{Review of Retrospective Cost Adaptive Control} \label{subsec:rcac}

RCAC is described in detail in \cite{rahmanCSM2017}. In this subsection, we summarize the main elements of this method.
Consider the strictly proper, discrete-time, input-output controller  
\begin{align}
    u_{\rmc, k} = \sum_{i=1}^{l_\rmc}P_{i,k}u_{k-i} + \sum_{i=1}^{l_\rmc}Q_{i,k}z_{k-i}, \label{IO_controller}
\end{align}
where $u_{\rmc, k} \in \mathbb R^{l_u}$ is the commanded input and the controller output $u_k \in \mathbb R^{l_u}$ is the control input,  $z_k \in \mathbb R^{l_z}$ is the measured performance variable, $l_\rmc$ is the controller-window length, and, for all $i\in \{1,\ldots,l_\rmc\},$  $P_{i,k} \in \mathbb R^{l_u \times l_u}$ and $Q_{i,k} \in \mathbb R^{l_u \times l_z}$ are the controller coefficient matrices.
Note that  $u_k$ results from applying constraints to $u_{\rmc, k},$ as shown in Subsection \ref{subsec:ctrl_impl}.
The controller \eqref{IO_controller} can be written as
\begin{align}
    u_{\rmc, k}   =   \phi_k  \theta_k , \label{controller}
\end{align}
where 
\begin{align}
	\phi_k &\isdef
    \left[ \arraycolsep=3pt\def\arraystretch{0.9} \begin{array}{cccccc} 
    			u_{k-1}^\rmT & \cdots & u_{k-l_\rmc}^\rmT & z_{k-1}^\rmT & \cdots & z_{k-l_\rmc}^\rmT
    \end{array} \right]
    		\otimes
    		I_{l_u}
    		\in \mathbb{R}^{l_u \times l_{\theta}},  \label{controller_phi} \\
    	\theta_k &\isdef {\rm vec}
    \left[ \arraycolsep=1.7pt\def\arraystretch{0.9} \begin{array}{cccccc} 
        P_{1,k} &\cdots &P_{l_\rmc,k} &Q_{1,k} &\cdots &Q_{l_\rmc,k}
    \end{array} \right] \in \BBR^{l_\theta},
\end{align}
$l_\theta \isdef l_\rmc l_u (l_u + l_z),$ and $\theta_k$ is the vector of controller coefficients, which are updated at each time step $k$.
If $z_k$ and $u_k$ are scalar, then the SISO transfer function of  \eqref{IO_controller} from $z_k$ to $u_k$ is given by
\begin{align}
G_{\rmc,k}(\bfq)= \frac{Q_{1,k} \bfq^{l_\rmc - 1} + \cdots + Q_{l_\rmc,k} }{\bfq^{l_\rmc} - P_{1,k}\bfq^{l_\rmc - 1} - \cdots - P_{l_\rmc,k} }.  
\end{align}

Next, define the retrospective cost variable
\begin{align}
	\hat z_k (\hat \theta) \isdef z_k  - G_\rmf(\textbf{q})(u_k - \phi_k \hat{\theta}), \label{zhat1}
\end{align}
where 
$G_{\rmf}$ is an $l_z \times l_u$ asymptotically stable, strictly proper transfer function, and
$\hat{\theta} \in \mathbb R^{l_\theta}$ is the controller coefficient vector determined by  optimization below.
The rationale underlying \eqref{zhat1} is to replace the applied past control inputs with the re-optimized control input $\phi_k \hat{\theta}$ 
so that the 
closed-loop transfer function from $u_k - \phi_k \theta_{k+1}$ to $z_k$
matches $G_{\rmf}$ \cite{rahmanCSM2017,islam2021data}.
Consequently, $G_{\rmf}$ serves as a closed-loop target model for  adaptation.

The filter $G_{\rmf}$ is called the {\it target model} and has the form
\begin{align}
	G_{\rmf}(\bfq) \isdef
			D_\rmf^{-1}(\bfq) N_\rmf(\bfq),
\end{align}
where $D_\rmf$ is an $l_z \times l_z$ polynomial matrix with leadin coefficient $I_{l_u},$ and $N_\rmf$ is an $l_z \times l_u$ polynomial matrix.
By defining the filtered variables $\phi_{\rmf, k} \in \BBR^{l_z \times l_\theta}$ and $u_{\rmf, k} \in \BBR^{l_z},$
\eqref{zhat1} can be written as 
\begin{align}
    \hat z_k(\hat \theta) = z_k - ( u_{\rmf, k} - \phi_{\rmf, k} \hat{\theta} ),
\end{align}
where
\begin{align}
    \phi_{\rmf, k} \isdef G_{\rmf} (\bfq) \phi_k, \quad u_{\rmf, k} \isdef G_{\rmf} (\bfq) u_k.
\end{align}

The choice of $G_\rmf$ includes all required modeling information.
When the plant is SISO, that is, $l_z=l_u=1,$ this information consists of the sign of the leading numerator coefficient, the relative degree of the sampled-data system, and all nonminimum-phase (NMP) zeros \cite{rahmanCSM2017,islam2021data}.
Since zeros are invariant under feedback, omission of a NMP zero from $G_\rmf$ may entail unstable pole-zero cancellation.
Cancellation can be prevented, however, by using the control weighting $R_u$ introduced below, as discussed in \cite{sumer2012retrospective,rahmanCSM2017}.
$G_\rmf$ can be constructed and updated online using data \cite{islam2021data}.
For simplicity in controlling the LP model, which is a SISO system, we fix $G_\rmf$ prior to implementation.

Using \eqref{zhat1}, we define the  cumulative cost function
\begin{align}
    J_k(\hat{\theta}) &\isdef \sum\limits_{i=0}^{k} [  \hat z_i^{\rm T}(\hat \theta) \hat z_i(\hat \theta)  +   (\phi_i \hat \theta)^{\rm T}  R_u \phi_i \hat \theta ] 
    +   (\hat \theta - \theta_0 ) ^{\rm T}   P_0^{-1} (\hat \theta - \theta_0 ), \label{eq:Jg}
\end{align}
%where $P_0 \in \BBR^{l_\theta \times l_\theta}$  and $R_u \in \BBR^{l_u \times l_u}$ are positive definite. 
%
where $P_0 \in \BBR^{l_\theta \times l_\theta}$ is positive definite, and $R_u \in \BBR^{l_u \times l_u}$ is positive semidefinite. 
As can be seen from \eqref{controller}, $R_u$ serves as a control weighting, which prevents RCAC from cancelling unmodeled NMP zeros, and the matrix $P_0^{-1}$ defines the regularization term and initializes the recursion for $P_k$ defined below.

The following result uses recursive least squares (RLS) 
\cite{ljung1983,islam2019recursive} to minimize \eqref{eq:Jg}, where, at each step $k,$  the minimizer of \eqref{eq:Jg} is the update $\theta_{k+1}$ of the controller coefficient vector.

\textit{Proposition 1}. For all $k \ge 0$, the unique global minimizer $\theta_{k+1}$ of \eqref{eq:Jg} is given by
\begin{align}
    P_{k+1}      &=  P_k  - P_k 
    \left[ \arraycolsep=0.9pt\def\arraystretch{0.9} \begin{array}{c} 
                \phi_{\rmf, k} \\
                \phi_k 
            \end{array}\right]^\rmT
    %\hspace{-0.5em} \Gamma_k^{-1}
    \hspace{-0.5em} \Gamma_k
    \left[ \arraycolsep=0.9pt\def\arraystretch{0.9} \begin{array}{c} 
                \phi_{\rmf, k} \\
                \phi_k 
            \end{array}\right]
    P_k , \label{eq:pk_update}\\
    \theta_{k+1} &= \theta_k  - P_{k+1}             
\left[ \arraycolsep=0.9pt\def\arraystretch{0.9} \begin{array}{c} 
                \phi_{\rmf, k} \\
                \phi_k 
            \end{array}\right]^\rmT \hspace{-0.5em} \bar{R}             
\left[ \arraycolsep=0.9pt\def\arraystretch{0.9} \begin{array}{cc} 
                 z_k - ( u_{\rmf, k} - \phi_{\rmf, k} \theta_k ) \\
                \phi_k\theta_k
            \end{array}\right] \label{eq:theta_update},
\end{align}
where
\begin{align}
   \Gamma_k &\isdef
                 \bar{R} - 
                    \bar{R}
                    \left[ \arraycolsep=1.1pt\def\arraystretch{0.9} \begin{array}{c} 
                \phi_{\rmf, k} \\
                \phi_k 
            \end{array}\right]
            %P_k
            \left(P_k^{-1} +
            \left[ \arraycolsep=1.1pt\def\arraystretch{0.9} \begin{array}{c} 
                \phi_{\rmf, k} \\
                \phi_k 
            \end{array}\right]^\rmT 
            \bar{R}
            \left[ \arraycolsep=1.1pt\def\arraystretch{0.9} \begin{array}{c} 
                \phi_{\rmf, k} \\
                \phi_k 
            \end{array}\right]
            \right)^{-1}
            \left[ \arraycolsep=1.1pt\def\arraystretch{0.9} \begin{array}{c} 
                \phi_{\rmf, k} \\
                \phi_k 
            \end{array}\right]^\rmT 
            \bar{R} 
            % \nonumber \\
            % &
            \in \BBR^{ (l_z + l_u) \times  (l_z + l_u) } ,\\
    \bar{R} &\isdef {\rm diag}( I_{l_z} , R_u ) \in \BBR^{ (l_z + l_u) \times  (l_z + l_u) }.
\end{align}
For all of the numerical simulations and physical experiments in this paper, $\theta_k$ is initialized as $\theta_0=0_{l_\theta\times 1}$ to reflect the absence of additional prior modeling information.
%
% Furthermore, except when specified otherwise, $G_\rmf(\textbf{q}) = -1/{\textbf{q}}$ ($N = -1$), where the minus sign reflects sign information and the relative degree is set to 1.
%
% Aside from the selection of hyperparameters discussed below, no other modeling information is used by RCAC.
%
For convenience, we set  $P_0 = p_0 I_{l_\theta},$ where the scalar $p_0>0$ determines the initial rate of adaptation.

\subsection{Target model $G_\rmf$ for systems with sinusoidal disturbances and time delays}

The structure of $G_\rmf$ can be designed to account for sinusoidal disturbances and time delays.
Note that time delays can also be induced by linear, time-invariant systems in applications that involve signals with sinusoidal components.
In this case, $G_\rmf$ is given by
\begin{equation}
    G_\rmf (\bfq) = \frac{N}{\bfq^{d_\rmf}} \left(\frac{1}{\bfq^2 - 2 \alpha_\rmf \cos(\omega_\rmf T_\rms) \bfq + \alpha_\rmf^2}\right)
\end{equation}
where $N \in \{-1, 1\}$ encodes the sign of the leading coefficient corresponding to the system model, $d_\rmf \ge 0$ is an integer that encodes the system time-delay, $\omega_\rmf \isdef 2 \pi f_\rmf,$ where $f_\rmf > 0$ encodes the disturbance frequency, and $\alpha_\rmf \in (0, 1]$ determines how close the eigenvalues of $G_\rmf$ are to the unit circle circumference.
In the case where $\alpha_\rmf = 1,$ the eigenvalues of $G_\rmf$ are placed at the unit circle circumference and for all $\alpha_\rmf < 1$ the eigenvalues are placed inside the unit circle.
While values of $\alpha_\rmf$ closer to 1 yield better distrubance rejection results, this can result in numerical instability.
Furthermore, the sign of $N$ can be inverted to account for a phase delay of $\pi$ rad, which can be used in cases where acceleration measurements are considered.

\subsection{Sampled-data Controller Implementation} \label{subsec:ctrl_impl}

\begin{figure} [h!]
    \centering
    \resizebox{0.6\columnwidth}{!}{%
    \begin{tikzpicture}[>={stealth'}, line width = 0.25mm]

    \node [input, name=ref]{};
    \node [sum, right = 0.75cm of ref, inner sep = 0.001em] (sum2) {};
    \node[draw = none] at (sum2.center) {$+$};
    \node [smallblock, rounded corners, right = 0.75cm of sum2 , minimum height = 0.6cm , minimum width = 0.7cm] (error_norm) {$\SF$};
    %
    %\draw [->] ([xshift=.25cm]error_norm.east)--([xshift=.25cm,yshift=-0.5cm]error_norm.east)--([xshift=0.6cm,yshift=-0.5cm]error_norm.east)--([xshift=1.4cm,yshift=0.5cm]error_norm.east);
    %
    \node [smallblock, rounded corners, right = 0.75cm of error_norm , minimum height = 1cm , minimum width = 0.7cm] (controller) {$G_{\rmc,k}$};
    \node [smallblock, rounded corners, right = 0.9cm of controller , minimum height = 0.6cm , minimum width = 0.7cm] (saturation) {$\sigma$};
    \node [smallblock, rounded corners, right = 0.75cm of saturation, minimum height = 0.6cm , minimum width = 0.5cm] (DA) {\scriptsize ZOH};
    \node [smallblock, fill=green!20, rounded corners, right = 0.75cm of DA, minimum height = 0.6cm , minimum width = 1cm] (system) {$\SM$};
    \node [output, right = 0.5cm of system] (output) {};
    \node [input, below = 1.25cm of system] (midpoint) {};
    
    \draw [->] (controller) -- node [above] {$u_{\rmc, k}$} (saturation);
    \draw [->] (saturation) -- node [above] {$u_k$} (DA);
    \draw [->] ([xshift = 0.3cm]saturation.east) |- ([xshift = -0.75cm, yshift = -0.75cm]controller.center) |- ([yshift = -0.3cm]controller.west);
    \draw [->] (DA) -- node [above] (du) {$u$} (system);

     %%%%%%%%%%%
    \node[circle,draw=black, fill=white, inner sep=0pt,minimum size=3pt] (rc11) at ([xshift=-2.9cm]midpoint) {};
    \node[circle,draw=black, fill=white, inner sep=0pt,minimum size=3pt] (rc21) at ([xshift=-3.2cm]midpoint) {};
    \draw [-] (rc21.north east) --node[below,yshift=.55cm]{$T_\rms$} ([xshift=.3cm,yshift=.15cm]rc21.north east) {};
    %%%%%%%%%%%
    
    \draw [->] (system) -- node [name=y, near end]{} node [very near end, above] {$y$}(output);
    
    \draw [->] (ref.east) -- node [near start, above,xshift = 0.1cm] {$r_k$}  node [near end, above] {} (sum2);
    \draw [-] (y.west) |- (midpoint);
    \draw [-] (midpoint) -| node [very near end, above, xshift=-0.7cm] {$y_k$} (rc11.east);
    \draw [->] (rc21) -| node [very near end, xshift = -0.35cm, yshift = 0.1cm] {\huge -} (sum2.south);
    \draw [->] (sum2.east) -- node [above, xshift = -0.05cm] {$e_k$} (error_norm.west);
    \draw [->] (error_norm.east) -- node [above, xshift = -0.05cm] {$z_k$} (controller.west);
    
    \end{tikzpicture}
    }  
    \caption{Sampled-data implementation of adaptive controller for control of the continuous-time system $\SM.$ 
    For this work, $r \equiv 0$ reflects the desire to minimize the oscillations in the measured signal, $\SF$ is a filter for signal conditioning, $\sigma$ is the control-magnitude saturation, and $\SM$ represents the LP model introduced in Section \ref{sec:LPM}.
    The form of filter $\SF$ depends on whether the available measurement $y$ is a displacement or an acceleration, and more details are given in Subsection \ref{subsec:filter}.
    }
    \label{AC_CT_blk_diag}
\end{figure}
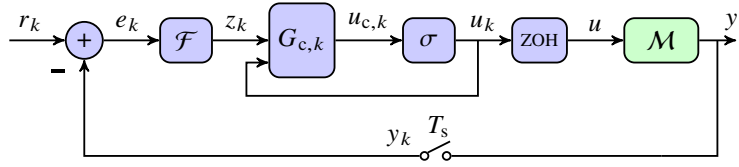

The adaptive controller is implemented as a sampled-data controller. 
Figure \ref{AC_CT_blk_diag} shows a block diagram of the sampled-data closed-loop system, where $y \in \BBR$ is the output of the continuous-time system $\SM$, $y_k$ is the sampled output, $r_k \in \BBR$ is the discrete-time command,  $e_k \isdef r_k - y_k$ is the command-following error, and $T_\rms>0$ is the controller sampling period.
%
%For all adaptive controller experiments, $T_\rms = 0.001$ s/step. 
%
%
%
The digital-to-analog (D/A) and analog-to-digital (A/D) interfaces, which are synchronous, are zero-order-hold (ZOH) and sampler, respectively.
For this work, $r \equiv 0$ reflects the desire to suppress oscillations in the measured signal.
Finally, $\SM$ represents the LP model introduced in Section \ref{sec:LPM}.

The measured performance variable $z_k$, which is used for adaptation, is the output of filter $\SF,$ which is used for signal conditioning.
The form of filter $\SF$ depends on whether the available measurement $y$ is a displacement or an acceleration, and more details are provided in Subsection \ref{subsec:filter}.
The adaptive controller $G_{\rmc, k}$ operates on $z_k$ to produce the discrete-time control $u_{\rmc, k} \in \BBR.$ 
Hence, $l_u = l_z = 1.$
Since the response of a real actuator is subjected to hardware constraints, the implemented discrete-time control is 
 $u_k\isdef \sigma(u_{\rmc,k}), $ 
where $\sigma\colon\BBR\to\BBR$ is the control-magnitude saturation function  
\begin{equation}
    \sigma(u)\isdef \begin{cases} u_{\max},&u> u_{\max},\\
    u,& u_{\min}\le u \le u_{\max},\\
    u_{\min}, & u<u_{\min},\end{cases}
\end{equation}
where $u_{\min},u_{\max}\in\BBR$ are the lower and upper magnitude saturation levels, respectively.
$G_{\rmc, k}, u_{\rmc, k},$ and $u_k$ are updated at each sampling time $t_k\isdef kT_\rms$. 
Implementation of the adaptive controller requires selection of the hyperparameters $l_\rmc$, $p_0,$ and $R_u$ depending on the system and performance requirements.
Note that $R_u$ is scalar since $l_u = 1.$

\subsection{Filter for Signal Conditioning} \label{subsec:filter}

The signal conditioning filter $\SF$ introduced in Subsection \ref{subsec:ctrl_impl} is used for improved suppression performance and its form depends on whether the available measurement is a displacement or an acceleration.
Three filters are considered depending on the available measurements.

In the case where displacement measurements are available, $\SF$ takes the form of a gain since displacement measurements in vibrational systems are small in magnitude; the filter is given by 
\begin{equation}
    \SF = K_\rmg,
\end{equation}
where $K_\rmg > 0.$

\begin{figure}[!ht]
\centering
\resizebox{0.8\columnwidth}{!}{%
\begin{tikzpicture}[>={stealth'}, line width = 0.25mm]
\node[draw = none] at (0,0) (orig) {};
\node [smallblock, rounded corners, minimum height = 3.25cm, minimum width = 14.75cm] at ([yshift = -1em, xshift = 17.3em]orig.center) (esc_controller) {};
\node[below right] at (esc_controller.north west) {\large$\SF$};
\node[smallblock, fill=red!20, rounded corners, minimum height = 3em, minimum width = 3em] at (orig.center) (HP_acc) {\Large$\frac{\bfq - 1}{\bfq - \frac{\nu_{\rm hp}}{\nu_{\rm hp} + 1} }$};
\node[below] at (HP_acc.south) {$\begin{array}{c} \mbox{High-Pass} \\ \mbox{Filter} \end{array}$};
\node[smallblock, fill=red!20, rounded corners, minimum height = 3em, minimum width = 3em, right = 3em of HP_acc.east] (trapz_int_acc) {\Large$\frac{T_\rms}{2} \frac{\bfq + 1}{\bfq - 1}$};
\node[below] at (trapz_int_acc.south) {$\begin{array}{c} \mbox{Trapezoidal} \\ \mbox{Integrator} \end{array}$};
\node[smallblock, fill=red!20, rounded corners, minimum height = 3em, minimum width = 3em, right = 3em of trapz_int_acc.east] (HP_vel) {\Large$\frac{\bfq - 1}{\bfq - \frac{\nu_{\rm hp}}{\nu_{\rm hp} + 1} }$};
\node[below] at (HP_vel.south) {$\begin{array}{c} \mbox{High-Pass} \\ \mbox{Filter} \end{array}$};
\node[smallblock, fill=red!20, rounded corners, minimum height = 3em, minimum width = 3em, right = 3em of HP_vel.east] (trapz_int_vel) {\Large$\frac{T_\rms}{2} \frac{\bfq + 1}{\bfq - 1}$};
\node[below] at (trapz_int_vel.south) {$\begin{array}{c} \mbox{Trapezoidal} \\ \mbox{Integrator} \end{array}$};
\node[smallblock, fill=red!20, rounded corners, minimum height = 3em, minimum width = 3em, right = 3em of trapz_int_vel.east] (HP_pos) {\Large$\frac{\bfq - 1}{\bfq - \frac{\nu_{\rm hp}}{\nu_{\rm hp} + 1} }$};
\node[below] at (HP_pos.south) {$\begin{array}{c} \mbox{High-Pass} \\ \mbox{Filter} \end{array}$};
\node[smallblock, fill=red!20, rounded corners, minimum height = 3em, minimum width = 3em, right = 3em of HP_pos.east] (gain) {\Large$K_\rmg$};
\node[below] at (gain.south) {Gain};
\draw [->] ([xshift = -3.5em]HP_acc.west) -- node [above, near start] {$e_k$} (HP_acc.west);
\draw [->] (HP_acc.east) -- (trapz_int_acc.west);
\draw [->] (trapz_int_acc.east) -- (HP_vel.west);
\draw [->] (HP_vel.east) -- (trapz_int_vel.west);
\draw [->] (trapz_int_vel.east) -- (HP_pos.west);
\draw [->] (HP_pos.east) -- (gain.west);
\draw [->] (gain.east) -- node [above, near end, xshift = -0.2em] {$z_k$} ([xshift = 3.5em]gain.east);
\end{tikzpicture}
}
\caption{
Signal conditioning filter $\SF$ in which the input $e_k$ is an acceleration measurement and the output $z_k$ is a displacement estimate.
$K_\rmg > 0$ is the filter gain, and $\nu_{\rm hp} > 0$ determines the high-pass filter cutoff frequency.
}
\label{fig:displacement_estimator}
\end{figure}
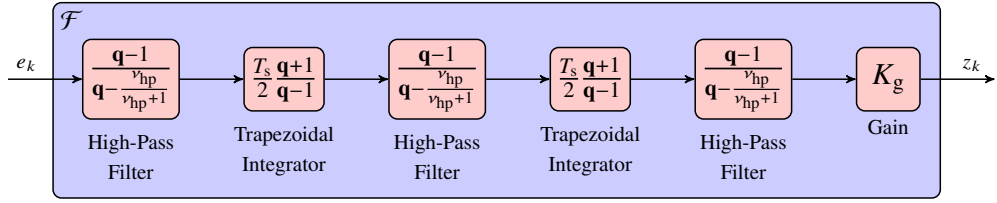

In the case where only acceleration measurements are available, we consider two filters with the objective of minimizing the magnitude of higher frequencies in the four acceleration signals that may be excited by the controller.
The first filter is a {\it low-pass filter} that yields a filtered acceleration signal, and is given by
\begin{equation}
    \SF = \frac{K_\rmg b_{\rm lp} \bfq}{\bfq^2 + a_{{\rm lp},1} \bfq + a_{{\rm lp},2}},
\end{equation}
where
\begin{align}
    a_{{\rm lp}, 1} &= -2 r_{\rm lp} \cos \theta_{\rm lp}, \\
    a_{{\rm lp}, 2} &= r_{\rm lp}^2, \\
    b_{\rm lp} &= 1 + a_{{\rm lp}, 1} + a_{{\rm lp}, 2}, \\
    r_{\rm lp} &= e^{- \zeta_{\rm lp} \omega_{\rm lp} T_\rms},\\
    \theta_{\rm lp} &= \omega_{\rm lp} T_\rms \sqrt{1 - \zeta_{\rm lp}^2},
\end{align}
$K_\rmg > 0$ is the filter gain, $\omega_{\rm lp} > 0$ is the filter cutoff frequency, and $\zeta_{\rm lp} \in (0, 1)$ is the filter damping.
The second filter is a {\it displacement estimation filter} composed by cascaded high-pass filters and integrators, based on the technique shown in \cite{zheng2019}, that yield a pseudo-displacement estimate signal, and is shown in Figure \ref{fig:displacement_estimator}, where $K_\rmg > 0$ is the filter gain, and $\nu_{\rm hp} > 0$ determines the high-pass filter cutoff frequency.
Note that the high-pass filter is equivalent to the subtraction of the original signal minus the cumulative average of the signal, calculated using a window of length $\nu_{\rm hp}.$

\section{Numerical Simulations} \label{sec:simulations}

In this section, we present the results of numerical simulations in which the RCAC algorithm is implemented to decrease the amplitude of the oscillations at a measurement point on the cantilever beam, caused by an external disturbance.
The results for feedback control using displacement measurements and acceleration measurements with a low-pass filter and a displacement estimation filter are shown.
In all cases, the open-loop results (no control) are compared to the closed-loop results (with feedback control), and RCAC is enabled at $t = 2.5$ s to allow the measured oscillations propagated by the disturbance enough time to settle.

For all numerical simulations, the simulation sampling period is $T_{\rm sim} = 10^{-4}$ s and the controller sampling period is $T_\rms = 2.5 \times 10^{-3}$ s.
The \texttt{ode45} solver from Matlab is used to integrate the model dynamics in between controller time steps.

The following cantilever system parameters are used:
\begin{equation*}
    L = 0.5 \ \rmm, \quad h = 0.05 \ \rmm, \quad b = 0.001 \ \rmm,  \quad m = 0.07 \mbox{ kg}, \quad E = 69 \times 10^{9} \mbox{ N/m}^2, \ \alpha = 1.5 \mbox{ s}^{-1}, \quad \beta = 2.5 \times 10^{-4} \mbox{ s},\\
\end{equation*}
\begin{equation*}
    %
    % n_\rmb = 20, \quad i_\rmd = 5, \quad i_\rmu = 12, \quad i_\rmy = 20.
    %
    n_\rmb = 20, \quad i_\rmd = 5, \quad i_\rmy = 20, \quad u_{\rm max} = -u_{\rm min} = 2.5,
\end{equation*}
where Young's modulus $E$ is chosen considering aluminum as the material of the beam,
and $\alpha$ and $\beta$ are chosen so that the minimum damping ratio corresponding to the eigenvalues of $A$ is approximately 0.1.
It follows from the value of $i_\rmd$ and $i_\rmy$ that the disturbance signal is applied at the fourth of the beam near the clamped end, and the measurement signal is obtained from the unclamped end of the beam.
Furthermore, for all $t\ge0,$ the disturbance signal is given by $d(t) = \sin (2 \pi \ f_{\rm dist} t),$ that is, a sinusoidal signal with unit amplitude and a frequency of $f_{\rm dist}$ Hz.

To test the effectiveness of RCAC under system parameter variations, the placement of the input $i_\rmu$ and the disturbance frequency $f_{\rm dist}$ are varied in different simulation runs, such that RCAC is tested under all combinations of $i_\rmu \in \{10, 12, 14, 16\}$ and $f_{\rm dist} = \{20 , 40, 60, 80\}$ Hz.
All tests are performed for each of the following cases:
\begin{enumerate}
    \item Displacement measurement feedback ($\SC_{\rm disp}$).
    \item Acceleration measurement feedback with low-pass filter ($\SC_{{\rm acc}, {\rm lp}}$).
    \item Acceleration measurement feedback with displacement estimation filter ($\SC_{{\rm acc}, {\rm disp \ est}}$).
\end{enumerate}
For each of these cases, a set of hyperparameters is fixed for all simulation runs, and $d_\rmf$ and $R_u$ are chosen for each $i_\rmu$ and $f_{\rm dist}$ combination to achieve the largest vibration attenuation, as these have the greatest impact on vibration attenuation performance.
For all cases, the hyperparameters of the target model $G_\rmf$ other than $d_\rmf$ and $N$ are fixed to $\omega_\rmf = 2 \pi f_\rmf = 2 \pi f_{\rm dist}$ and $\alpha_\rmf = 0.95.$

Furthermore, to quantify vibration attenuation performance, 
% the displacement amplitude decrease $y_{{\rm disp}, {\rm dec}}$ is defined as
we define
\begin{equation}
    y_{{\rm disp}, {\rm dec}} \isdef \frac{ y_{{\rm disp}, {\rm OL}}}{ y_{{\rm disp}, {\rm CL}}}, \label{eq:amp_decrease}
\end{equation}
where $y_{{\rm disp}, {\rm OL}}$ is the steady-state amplitude of the displacement at $i_\rmy$ in the open-loop case and $y_{{\rm disp}, {\rm CL}}$ is the is the steady-state amplitude of the displacement at $i_\rmy$ in the closed-loop case.
The simulations are run until the displacement amplitudes converge to a constant value; $y_{{\rm disp}, {\rm OL}}$ and $y_{{\rm disp}, {\rm CL}}$ are obtained as the maximum measured displacement amplitudes of the last 0.5 s of the simulation time.

\subsection{Displacement Measurements}\label{subsec:disp_results}

\begin{figure}[h!]
    \centering
    \includegraphics[width = 0.75\columnwidth]{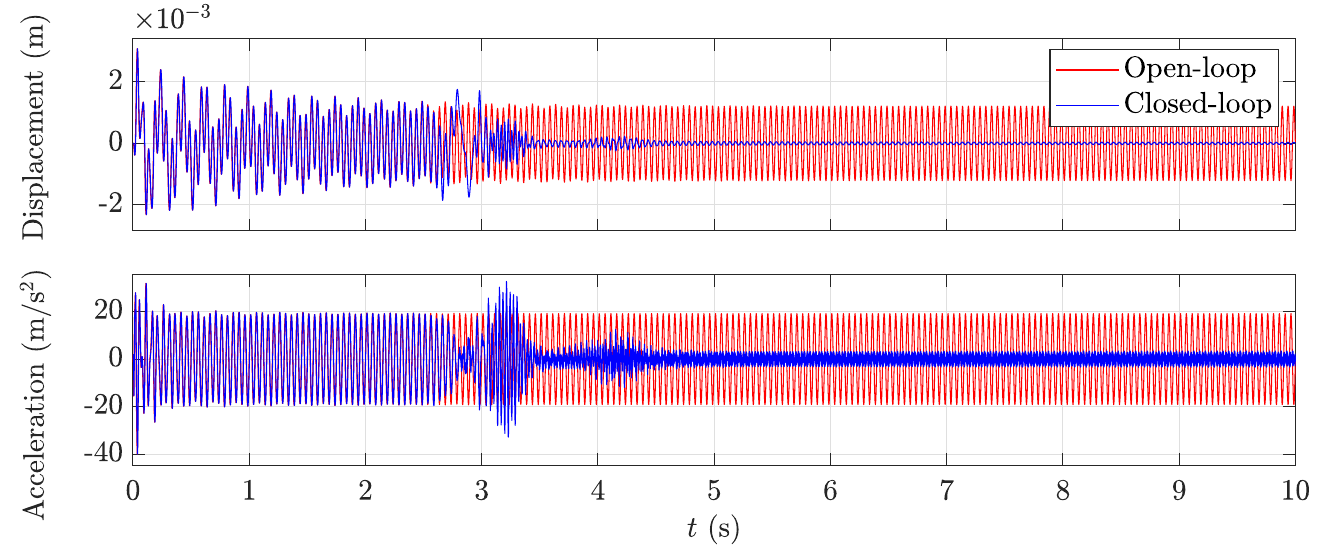}
    \caption{Subsection \ref{subsec:disp_results}: Open-loop and closed-loop displacement and acceleration results in the time domain with RCAC using displacement measurements $\SC_{\rm disp}$ in the case where $i_\rmu = 12$ and $f_{\rm dist} = 20$ Hz with $d_\rmf = 5$ and $R_u = 1.$}
    \label{fig:pos_fb_time}
\end{figure}

\begin{figure}[h!]
    \centering
    \includegraphics[width = 0.75\columnwidth]{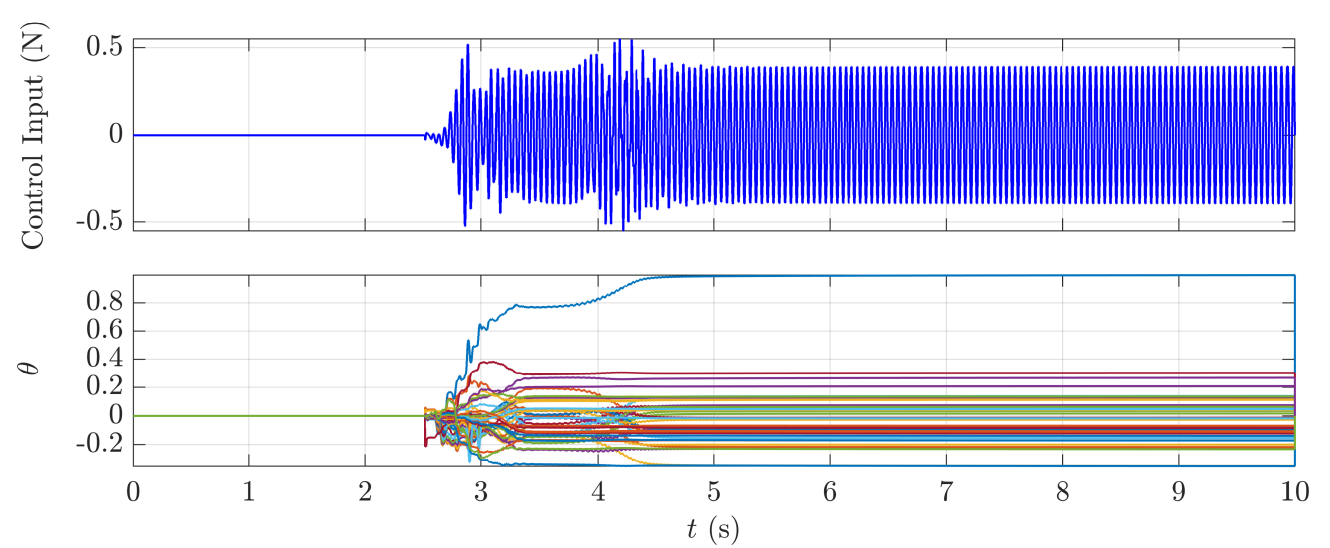}
    \caption{Subsection \ref{subsec:disp_results}: Control input and RCAC coefficients $\theta$ in the time domain with RCAC using displacement measurements $\SC_{\rm disp}$ in the case where $i_\rmu = 12$ and $f_{\rm dist} = 20$ Hz with $d_\rmf = 5$ and $R_u = 1.$}
    \label{fig:pos_fb_ctrl}
\end{figure}

\begin{figure}[h!]
    \centering
    \includegraphics[width = 0.75\columnwidth]{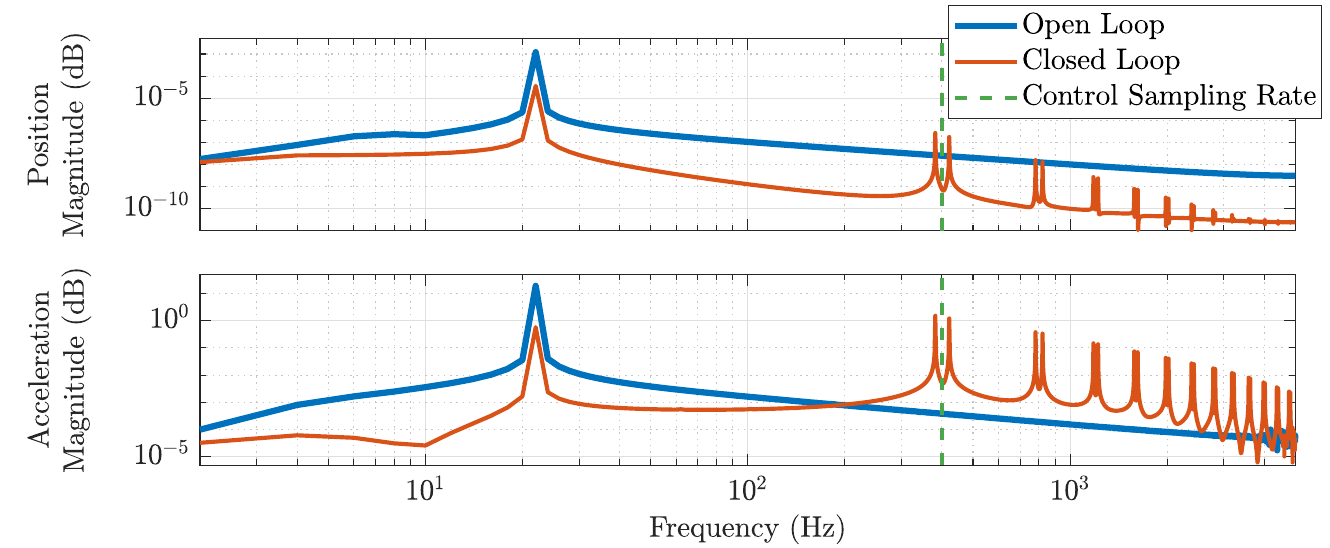}
    \caption{Subsection \ref{subsec:disp_results}: Open-loop and closed-loop displacement and acceleration results in the frequency domain with RCAC using displacement measurements $\SC_{\rm disp}$ in the case where $i_\rmu = 12$ and $f_{\rm dist} = 20$ Hz with $d_\rmf = 5$ and $R_u = 1.$}
    \label{fig:pos_fb_freq}
\end{figure}

For RCAC implementation using displacement measurement feedback $\SC_{\rm disp}$, the following hyperparameters are fixed for all simulation tests:
\begin{equation*}
    l_\rmc = 20, \quad p_0 = 1, \quad N = -1, \quad u_{\rm max} = -u_{\rm min} = 2.5, \quad K_\rmg = 500.
\end{equation*}
To illustrate the effectiveness of RCAC for vibration suppression, the results of this implementation in the case where $i_\rmu = 12$ and $f_{\rm dist} = 20$ Hz with $d_\rmf = 5$ and $R_u = 1$ are shown in Figures \ref{fig:pos_fb_time}, \ref{fig:pos_fb_ctrl}, and \ref{fig:pos_fb_freq}.
In particular, it follows from Figure \ref{fig:pos_fb_time} that the resulting vibration attenuation is given by $y_{{\rm disp}, {\rm dec}} = 30.38$ dB.
Finally, the vibration attenuation obtained in simulation runs under all combinations of $i_\rmu \in \{10, 12, 14, 16\}$ and $f_{\rm dist} = \{20 , 40, 60, 80\}$ Hz is summarized in Table \ref{tab:ex_disp_results}, which also include the choice of $d_\rmf$ and $R_u$ used to obtain each result.

\begin{table}[h!]
\caption{Subsection \ref{subsec:disp_results}: Vibration attenuation results with RCAC using displacement measurements $\SC_{\rm disp}$ for all combinations of $i_\rmu \in \{10, 12, 14, 16\}$ and $f_{\rm dist} = \{20 , 40, 60, 80\}$ Hz. The table displays the vibration attenuation $y_{{\rm disp}, {\rm dec}}$ defined in \eqref{eq:amp_decrease}, and the hyperparameters $d_\rmf$ and $R_u$ used to obtain these results.}
\label{tab:ex_disp_results}
\centering
\renewcommand{\arraystretch}{1.4}
\resizebox{0.7\columnwidth}{!}{%
\begin{tabular}{l !{\vrule width 0.6mm} c !{\vrule width 0.1mm} c !{\vrule width 0.1mm} c !{\vrule width 0.1mm} c !{\vrule width 0.6mm} c !{\vrule width 0.1mm} c !{\vrule width 0.1mm} c !{\vrule width 0.1mm} c !{\vrule width 0.6mm} c !{\vrule width 0.1mm} c !{\vrule width 0.1mm} c !{\vrule width 0.1mm} c !{\vrule width 0.6mm} }
\clineB{2-13}{4}
&\multicolumn{4}{c!{\vrule width 0.6mm} }{Vibration attenuation $y_{{\rm disp}, {\rm dec}}$ (dB)} & \multicolumn{4}{c!{\vrule width 0.6mm} }{$d_\rmf$} & \multicolumn{4}{c!{\vrule width 0.6mm} }{$R_u$} \\
\nhline{0.6mm}
\multicolumn{1}{!{\vrule width 0.6mm} l !{\vrule width 0.6mm}}{\diagbox[trim=r,height = 2\line, width = 4em,linewidth=0.6mm, outerrightsep=-6pt, innerrightsep=1pt, innerleftsep=2pt]%
    {$f_{\rm dist}$}{\hspace{-1em}$i_\rmu$}} & 10 & 12 & 14 & 16 & 10 & 12 & 14 & 16 & 10 & 12 & 14 & 16 \\
\nhline{0.6mm}
\multicolumn{1}{!{\vrule width 0.6mm} l !{\vrule width 0.6mm}}{20 Hz} & 30.97 & 30.38 & 31.67 & 32.41 & 5 & 5 & 5 & 5 & 1 & 1 & 1 & 1 \\
\nhline{0.1mm}
\multicolumn{1}{!{\vrule width 0.6mm} l !{\vrule width 0.6mm}}{40 Hz} & 28.78 & 27.88 & 4.16 & 26.97 & 9 & 8 & 7 & 3 & 0.5 & 0.4 & 0.5 & 0.4 \\
\nhline{0.1mm}
\multicolumn{1}{!{\vrule width 0.6mm} l !{\vrule width 0.6mm}}{60 Hz} & 28.51 & 2.01 & 21.37 & 29.62 & 6 & 5 & 9 & 3 & 0.05 & 0.1 & 0.1 & 0.05 \\
\nhline{0.1mm}
\multicolumn{1}{!{\vrule width 0.6mm} l !{\vrule width 0.6mm}}{80 Hz} & 8.30 & 21.39 & 28.00 & 27.34 & 9 & 7 & 7 & 7 & 0.1 & 0.05 & 0.05 & 0.05 \\
\nhline{0.6mm}
\end{tabular}
}
\end{table}

\subsection{Acceleration Measurements and Low-Pass Filter} \label{subsec:acc_lp_filt_results}

\begin{figure}[h!]
    \centering
    \includegraphics[width = 0.75\columnwidth]{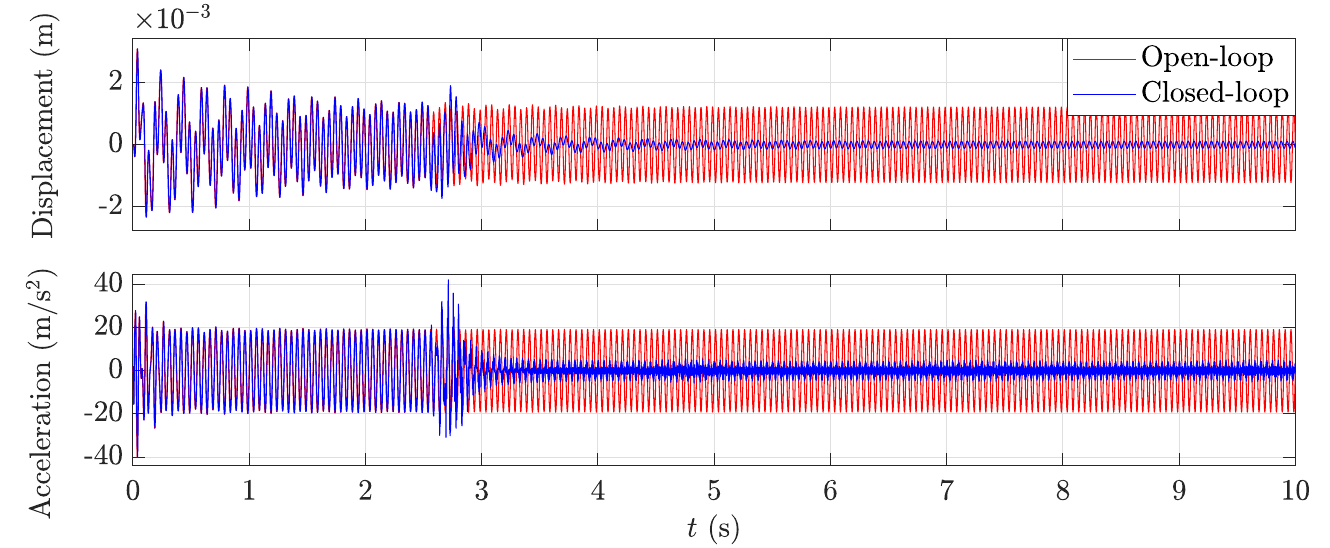}
    \vspace{-0.5em}
    \caption{Subsection \ref{subsec:acc_lp_filt_results}: Open-loop and closed-loop displacement and acceleration results in the time domain with RCAC using acceleration measurements with a low-pass filter $\SC_{{\rm acc}, {\rm lp}}$ in the case where $i_\rmu = 12$ and $f_{\rm dist} = 20$ Hz with $d_\rmf = 5$ and $R_u = 1.$}
    \label{fig:acc_fb_time}
\end{figure}

\begin{figure}[h!]
    \centering
    \includegraphics[width = 0.75\columnwidth]{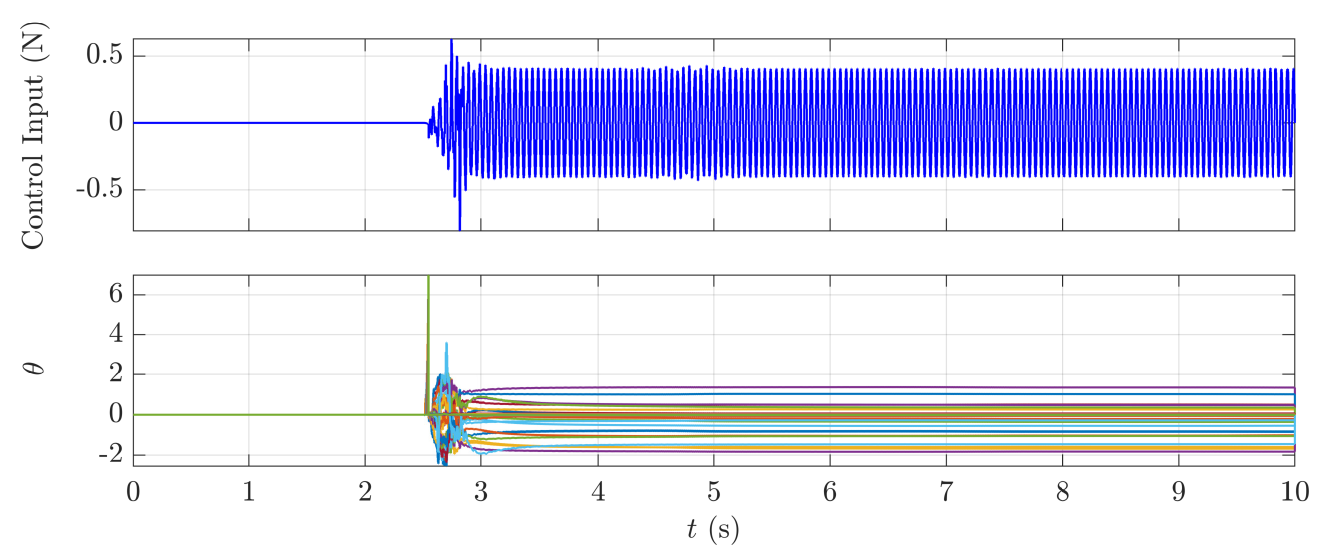}
    \caption{Subsection \ref{subsec:acc_lp_filt_results}: Control input and RCAC coefficients $\theta$ in the time domain with RCAC using acceleration measurements with a low-pass filter $\SC_{{\rm acc}, {\rm lp}}$ in the case where $i_\rmu = 12$ and $f_{\rm dist} = 20$ Hz with $d_\rmf = 5$ and $R_u = 1.$}
    \label{fig:acc_fb_ctrl}
\end{figure}

\begin{figure}[h!]
    \centering
    \includegraphics[width = 0.75\columnwidth]{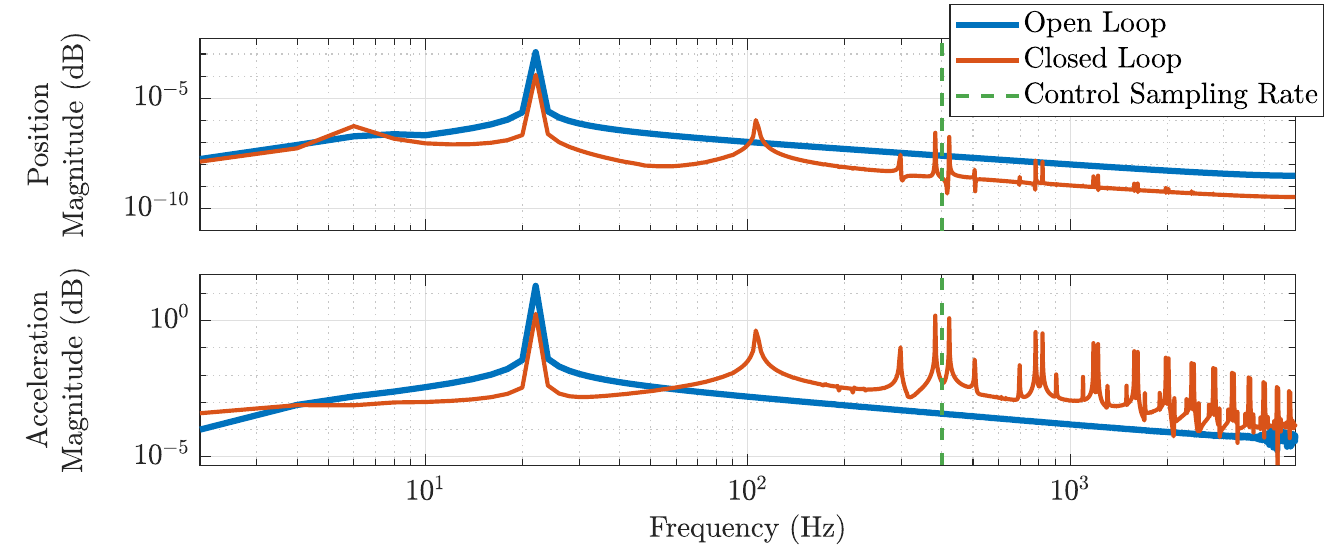}
    \caption{Subsection \ref{subsec:acc_lp_filt_results}: Open-loop and closed-loop displacement and acceleration results in the frequency domain with RCAC using acceleration measurements with a low-pass filter $\SC_{{\rm acc}, {\rm lp}}$ in the case where $i_\rmu = 12$ and $f_{\rm dist} = 20$ Hz with $d_\rmf = 5$ and $R_u = 1.$}
    \label{fig:acc_fb_freq}
\end{figure}

For RCAC implementation using acceleration measurement feedback with a low-pass filter $\SC_{{\rm acc}, {\rm lp}}$, the following hyperparameters are fixed for all simulation tests:
\begin{equation*}
    l_\rmc = 20, \quad p_0 = 1, \quad R_u = 40, \quad N = 1, \quad K_\rmg = 1, \quad \omega_{\rm lp} = 2 \pi \ 150 , \quad \zeta_{\rm lp} = 0.8.
\end{equation*}
To illustrate the effectiveness of RCAC for vibration suppression, the results of this implementation in the case where $i_\rmu = 12$ and $f_{\rm dist} = 20$ Hz with $d_\rmf = 5$ and $R_u = 1$ are shown in Figures  \ref{fig:acc_fb_time}, \ref{fig:acc_fb_ctrl}, and \ref{fig:acc_fb_freq}.
In particular, it follows from Figure \ref{fig:acc_fb_time}  that the resulting vibration attenuation is given by $y_{{\rm disp}, {\rm dec}} = 20.42$ dB.
Finally, the vibration attenuation obtained in simulation runs under all combinations of $i_\rmu \in \{10, 12, 14, 16\}$ and $f_{\rm dist} = \{20 , 40, 60, 80\}$ Hz is summarized in Table \ref{tab:ex_acc_lp_filt_results}, which also include the choice of $d_\rmf$ and $R_u$ used to obtain each result.

\begin{table}[h!]
\caption{Subsection \ref{subsec:acc_lp_filt_results}: Vibration attenuation results with RCAC using acceleration measurements with a low-pass filter $\SC_{{\rm acc}, {\rm lp}}$ for all combinations of $i_\rmu \in \{10, 12, 14, 16\}$ and $f_{\rm dist} = \{20 , 40, 60, 80\}$ Hz. The table displays the vibration attenuation $y_{{\rm disp}, {\rm dec}}$ defined in \eqref{eq:amp_decrease}, and the hyperparameters $d_\rmf$ and $R_u$ used to obtain these results.}
\label{tab:ex_acc_lp_filt_results}
\centering
\renewcommand{\arraystretch}{1.4}
\resizebox{0.7\columnwidth}{!}{%
\begin{tabular}{l !{\vrule width 0.6mm} c !{\vrule width 0.1mm} c !{\vrule width 0.1mm} c !{\vrule width 0.1mm} c !{\vrule width 0.6mm} c !{\vrule width 0.1mm} c !{\vrule width 0.1mm} c !{\vrule width 0.1mm} c !{\vrule width 0.6mm} c !{\vrule width 0.1mm} c !{\vrule width 0.1mm} c !{\vrule width 0.1mm} c !{\vrule width 0.6mm} }
\clineB{2-13}{4}
&\multicolumn{4}{c!{\vrule width 0.6mm} }{Vibration attenuation $y_{{\rm disp}, {\rm dec}}$ (dB)} & \multicolumn{4}{c!{\vrule width 0.6mm} }{$d_\rmf$} & \multicolumn{4}{c!{\vrule width 0.6mm} }{$R_u$} \\
\nhline{0.6mm}
\multicolumn{1}{!{\vrule width 0.6mm} l !{\vrule width 0.6mm}}{\diagbox[trim=r,height = 2\line, width = 4em,linewidth=0.6mm, outerrightsep=-6pt, innerrightsep=1pt, innerleftsep=2pt]%
    {$f_{\rm dist}$}{\hspace{-1em}$i_\rmu$}} & 10 & 12 & 14 & 16 & 10 & 12 & 14 & 16 & 10 & 12 & 14 & 16 \\
\nhline{0.6mm}
\multicolumn{1}{!{\vrule width 0.6mm} l !{\vrule width 0.6mm}}{20 Hz} & 30.27 & 20.42 & 16.07 & 22.81 & 3 & 3 & 3 & 3 & 40 & 40 & 30 & 30 \\
\nhline{0.1mm}
\multicolumn{1}{!{\vrule width 0.6mm} l !{\vrule width 0.6mm}}{40 Hz} & 24.20 & 23.09 & 5.43 & 26.83 & 8 & 7 & 7 & 8 & 40 & 40 & 100 & 40 \\
\nhline{0.1mm}
\multicolumn{1}{!{\vrule width 0.6mm} l !{\vrule width 0.6mm}}{60 Hz} & 21.67 & -2.37 & 6.49 & 26.14 & 9 & 20 & 20 & 18 & 60 & 1 & 50 & 50 \\
\nhline{0.1mm}
\multicolumn{1}{!{\vrule width 0.6mm} l !{\vrule width 0.6mm}}{80 Hz} & 8.79 & 10.80 & 14.47 & 26.92 & 20 & 19 & 19 & 19 & 40 & 50 & 100 & 60 \\

\nhline{0.6mm}
\end{tabular}
}
\end{table}

\subsection{Acceleration Measurements and Displacement Estimation Filter} \label{subsec:acc_displ_est_filt_results}

\begin{figure}[h!]
    \centering
    \includegraphics[width = 0.75\columnwidth]{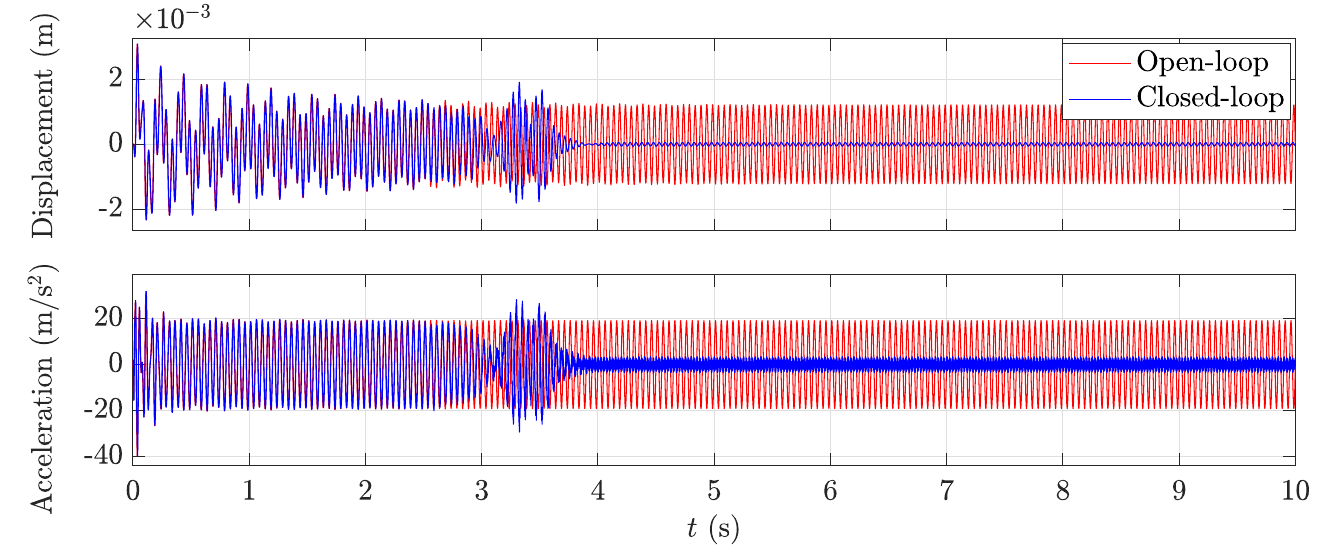}
    \vspace{-0.5em}
    \caption{Subsection \ref{subsec:acc_displ_est_filt_results}: Open-loop and closed-loop displacement and acceleration results in the time domain with RCAC using acceleration measurements with a displacement estimation filter $\SC_{{\rm acc}, {\rm disp \ est}}$ in the case where $i_\rmu = 12$ and $f_{\rm dist} = 20$ Hz with $d_\rmf = 5$ and $R_u = 1.$}
    \label{fig:pos_est_fb_time}
\end{figure}

\begin{figure}[h!]
    \centering
    \includegraphics[width = 0.75\columnwidth]{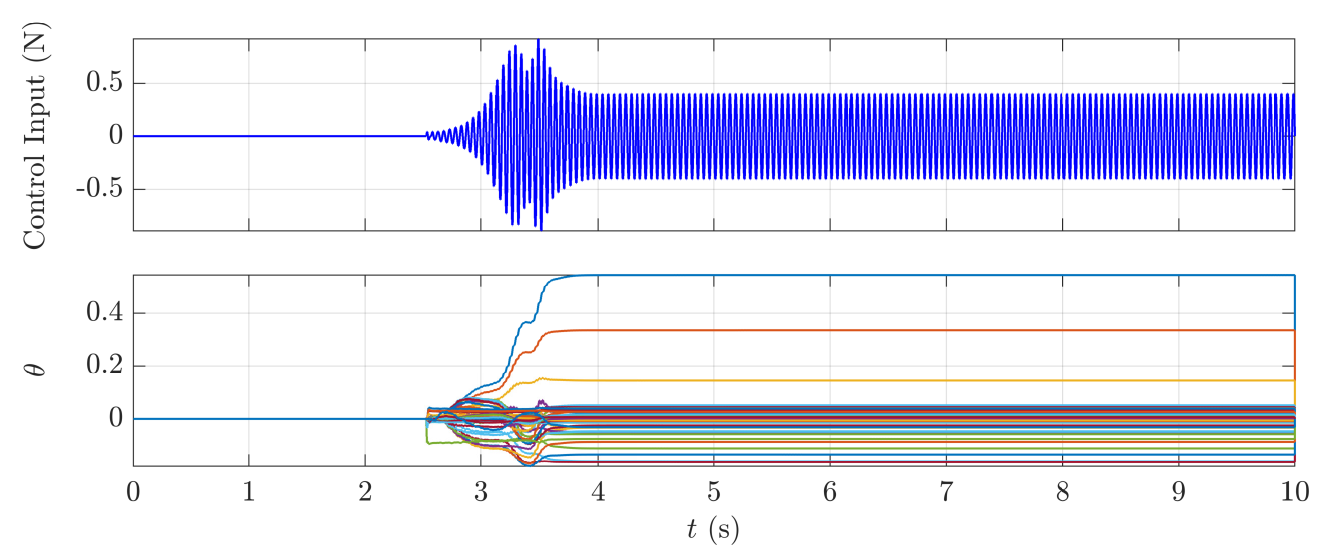}
    \caption{Subsection \ref{subsec:acc_displ_est_filt_results}: Control input and RCAC coefficients $\theta$ in the time domain with RCAC using acceleration measurements with a displacement estimation filter $\SC_{{\rm acc}, {\rm disp \ est}}$ in the case where $i_\rmu = 12$ and $f_{\rm dist} = 20$ Hz with $d_\rmf = 5$ and $R_u = 1.$}
    \label{fig:pos_est_fb_ctrl}
\end{figure}

\begin{figure}[h!]
    \centering
    \includegraphics[width = 0.75\columnwidth]{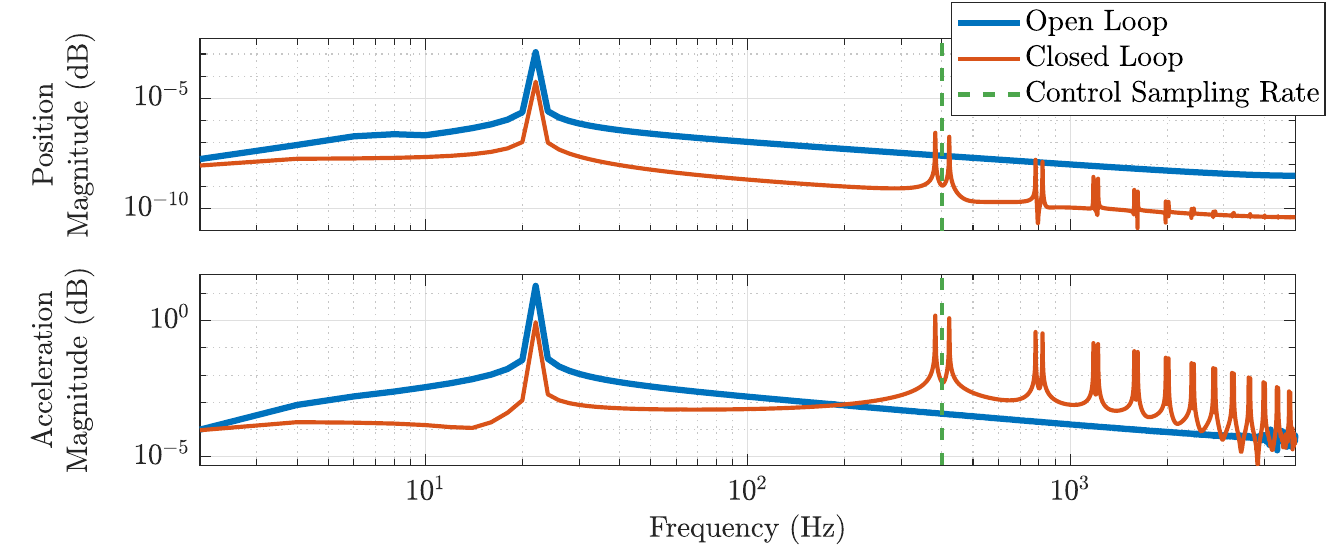}
    \caption{Subsection \ref{subsec:acc_displ_est_filt_results}: Open-loop and closed-loop displacement and acceleration results in the frequency domain with RCAC using acceleration measurements with a displacement estimation filter $\SC_{{\rm acc}, {\rm disp \ est}}$ in the case where $i_\rmu = 12$ and $f_{\rm dist} = 20$ Hz with $d_\rmf = 5$ and $R_u = 1.$}
    \label{fig:pos_est_fb_freq}
\end{figure}

For RCAC implementation using acceleration measurement feedback with a displacement estimation filter $\SC_{{\rm acc}, {\rm disp \ est}}$, the following hyperparameters are fixed for all simulation tests:
\begin{equation*}
    l_\rmc = 25, \quad p_0 = 0.1, \quad N = -1, \quad K_\rmg = 500, \quad \nu_{\rm hp} = 20.
\end{equation*}
To illustrate the effectiveness of RCAC for vibration suppression, the results of this implementation in the case where $i_\rmu = 12$ and $f_{\rm dist} = 20$ Hz with $d_\rmf = 8$ and $R_u = 1$ are shown in Figures \ref{fig:pos_est_fb_time}, \ref{fig:pos_est_fb_ctrl}, and \ref{fig:pos_est_fb_freq}.
In particular, it follows from Figure \ref{fig:pos_est_fb_time} that the resulting vibration attenuation is given by $y_{{\rm disp}, {\rm dec}} = 26.77$ dB.
Finally, the vibration attenuation obtained in simulation runs under all combinations of $i_\rmu \in \{10, 12, 14, 16\}$ and $f_{\rm dist} = \{20 , 40, 60, 80\}$ Hz is summarized in Table \ref{tab:ex_acc_displ_est_results}, which also include the choice of $d_\rmf$ and $R_u$ used to obtain each result.

\clearpage

\begin{table}[h!]
\caption{Subsection \ref{subsec:acc_displ_est_filt_results}: Vibration attenuation results with RCAC using acceleration measurements with a displacement estimation filter $\SC_{{\rm acc}, {\rm disp \ est}}$ for all combinations of $i_\rmu \in \{10, 12, 14, 16\}$ and $f_{\rm dist} = \{20 , 40, 60, 80\}$ Hz. The table displays the vibration attenuation $y_{{\rm disp}, {\rm dec}}$ defined in \eqref{eq:amp_decrease}, and the hyperparameters $d_\rmf$ and $R_u$ used to obtain these results.}
\label{tab:ex_acc_displ_est_results}
\centering
\renewcommand{\arraystretch}{1.4}
\resizebox{0.7\columnwidth}{!}{%
\begin{tabular}{l !{\vrule width 0.6mm} c !{\vrule width 0.1mm} c !{\vrule width 0.1mm} c !{\vrule width 0.1mm} c !{\vrule width 0.6mm} c !{\vrule width 0.1mm} c !{\vrule width 0.1mm} c !{\vrule width 0.1mm} c !{\vrule width 0.6mm} c !{\vrule width 0.1mm} c !{\vrule width 0.1mm} c !{\vrule width 0.1mm} c !{\vrule width 0.6mm} }
\clineB{2-13}{4}
&\multicolumn{4}{c!{\vrule width 0.6mm} }{Vibration attenuation $y_{{\rm disp}, {\rm dec}}$ (dB)} & \multicolumn{4}{c!{\vrule width 0.6mm} }{$d_\rmf$} & \multicolumn{4}{c!{\vrule width 0.6mm} }{$R_u$} \\
\nhline{0.6mm}
\multicolumn{1}{!{\vrule width 0.6mm} l !{\vrule width 0.6mm}}{\diagbox[trim=r,height = 2\line, width = 4em,linewidth=0.6mm, outerrightsep=-6pt, innerrightsep=1pt, innerleftsep=2pt]%
    {$f_{\rm dist}$}{\hspace{-1em}$i_\rmu$}} & 10 & 12 & 14 & 16 & 10 & 12 & 14 & 16 & 10 & 12 & 14 & 16 \\
\nhline{0.6mm}
\multicolumn{1}{!{\vrule width 0.6mm} l !{\vrule width 0.6mm}}{20 Hz} & 27.19 & 26.77 & 23.49 & 28.85 & 3 & 8 & 8 & 8 & 1 & 1 & 5 & 1 \\
\nhline{0.1mm}
\multicolumn{1}{!{\vrule width 0.6mm} l !{\vrule width 0.6mm}}{40 Hz} & 26.18 & 28.47 & 4.10 & 25.90 & 8 & 5 & 6 & 5 & 0.05 & 0.5 & 2 & 0.75 \\
\nhline{0.1mm}
\multicolumn{1}{!{\vrule width 0.6mm} l !{\vrule width 0.6mm}}{60 Hz} & 14.30 & 0.81 & 20.19 & 21.90 & 6 & 5 & 4 & 9 & 0.1 & 1 & 0.75 & 0.05 \\
\nhline{0.1mm}
\multicolumn{1}{!{\vrule width 0.6mm} l !{\vrule width 0.6mm}}{80 Hz} & 8.16 & 9.54 & 15.20 & 27.02 & 9 & 7 & 8 & 8 & 0.05 & 0.1 & 0.1 & 0.1 \\
\nhline{0.6mm}
\end{tabular}
}
\end{table}

\subsection{Comparison of closed-loop performance} \label{subsec:closed_loop_performance}

\begin{figure}[h!]
    \centering
    \includegraphics[width = 0.6\columnwidth]{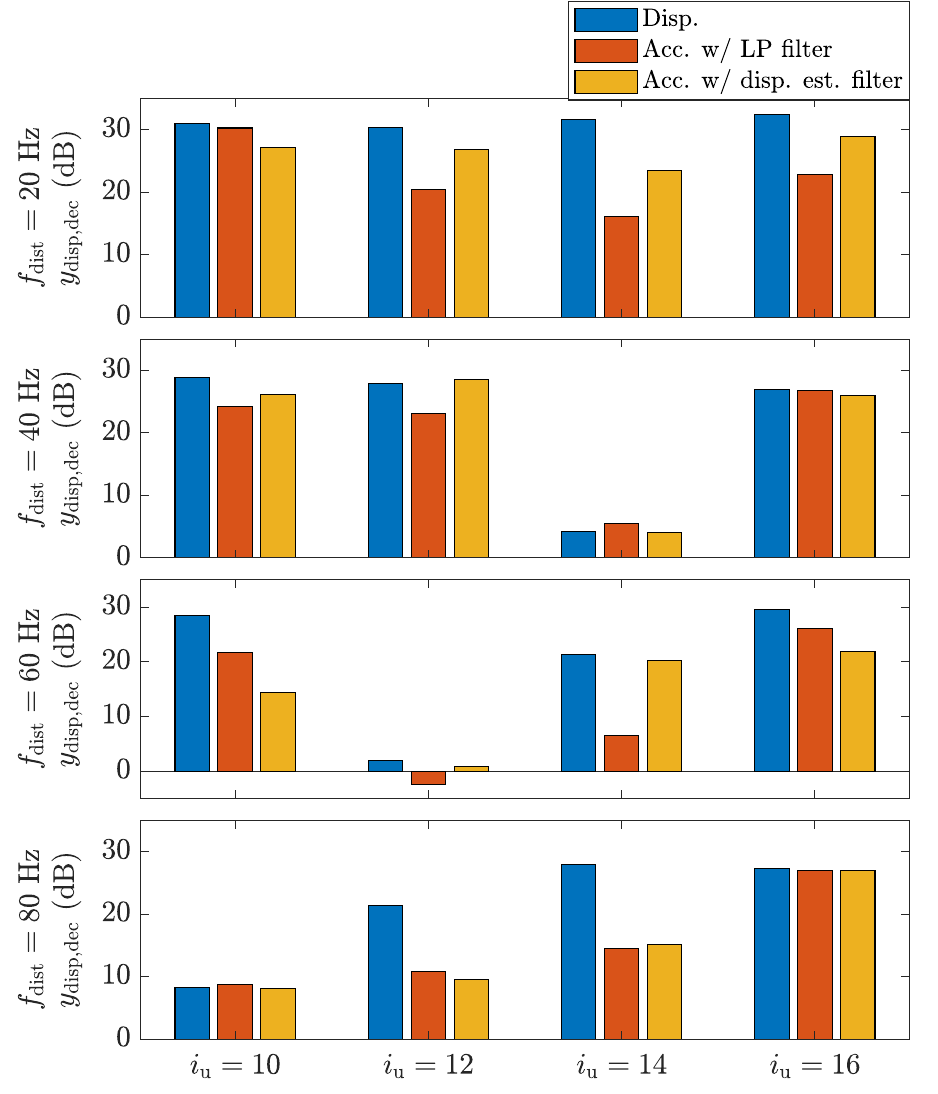}
    \caption{Subsection \ref{subsec:closed_loop_performance}: Vibration attenuation $y_{{\rm disp}, {\rm dec}}$  with RCAC using displacement measurements, acceleration measurements with a low-pass filter, and  acceleration measurements with a displacement estimation filter. The results are shown for all combinations of $i_\rmu \in \{10, 12, 14, 16\}$ and $f_{\rm dist} = \{20 , 40, 60, 80\}$ Hz.}
    \label{fig:bar_plot_comparison}
\end{figure}

The results of the three measurement and filter combinations ($\SC_{\rm disp}$, $\SC_{{\rm acc}, {\rm lp}}$, $\SC_{{\rm acc}, {\rm disp \ est}}$) are summarized and compared in Figure \ref{fig:bar_plot_comparison}.
In most cases, $\SC_{{\rm acc}, {\rm disp \ est}}$ improves upon $\SC_{{\rm acc}, {\rm lp}}$.
However, the performance of $\SC_{{\rm acc}, {\rm disp \ est}}$ deteriorates at higher frequencies ($f_{\rm dist} \in \{60, 80\}$ Hz).
Since displacement estimation filter hyperparameter $\nu_{\rm hp}$ was chosen considering the case in which $f_{\rm dist} = 20$ Hz, retuning $\nu_{\rm hp}$ may be required for improved performance at higher frequencies.
Hence, $\SC_{{\rm acc}, {\rm disp \ est}}$ offers the possibility of recovering the performance from displacement feedback $\SC_{\rm disp}$.

Another advantage of $\SC_{{\rm acc}, {\rm disp \ est}}$ is that it possesses similar closed-loop behavior to $\SC_{\rm disp}.$
To illustrate this, note that the acceleration magnitude frequency response of the closed-loop result corresponding to $\SC_{{\rm acc}, {\rm lp}},$ shown in Figure \ref{fig:acc_fb_freq}, shows two more frequency peaks at frequencies lower than the control sampling frequency than the frequency peaks in the  frequency responses corresponding to $\SC_{\rm disp}$ and $\SC_{{\rm acc}, {\rm disp \ est}}$, shown in Figures \ref{fig:pos_fb_freq} and \ref{fig:pos_est_fb_freq}.

Finally, it follows from Tables \ref{tab:ex_disp_results}, \ref{tab:ex_acc_lp_filt_results}, and \ref{tab:ex_acc_displ_est_results} that the optimal hyperparameters for $\SC_{{\rm acc}, {\rm disp \ est}}$ are closer to those corresponding to $\SC_{\rm disp}$ than those corresponding to $\SC_{{\rm acc}, {\rm lp}},$ especially in the cases where $f_{\rm dist} \in \{60, 80\}$ Hz.
Hence, the hyperparameter space considered for $\SC_{\rm disp}$ can also be used for $\SC_{{\rm acc}, {\rm disp \ est}}.$

\section{Conclusions} \label{sec:conclusions}

This paper presented a lumped-parameter model of a cantilever beam and applied the retrospective cost adaptive control (RCAC) algorithm for suppressing vibrations induced by external sinusoidal disturbances, without requiring any prior model information of the beam.
% without requiring any prior model information other than the frequency of the disturbance.
%
Numerical simulations demonstrate that RCAC can effectively attenuate vibration using both displacement and acceleration feedback.
%
% As observed in the numerical simulation results,
% Numerical results suggest that the displacement feedback yields better closed-loop performance than acceleration feedback. 
% %
% However, in practical applications, acceleration measurements are considerably easier to obtain than displacement measurements.
% %
% While acceleration can be derived by differentiating displacement twice, the inverse, recovering displacement by integrating acceleration twice, is generally infeasible due to unknown or imprecise initial conditions.
Simulation results indicate that displacement feedback provides superior closed-loop performance compared to acceleration feedback.
However, in practical applications, acceleration measurements are significantly easier to obtain and more reliable than displacement measurements.
While acceleration can be obtained directly from sensors or by differentiating displacement twice, the reverse process, that is, recovering displacement through double integration of acceleration, is generally infeasible due to uncertainty in the initial conditions.
%
% Nevertheless, in a vibrating system, since both acceleration and displacement remain bounded, a pseudo-displacement estimate can be obtained by conditioning the acceleration measurement as discussed in Section \ref{subsec:filter}.
To address this limitation, a pseudo-displacement estimate was formulated by conditioning the acceleration measurements, as discussed in Section \ref{subsec:filter}.
%
% While the simulation results show that the pseudo-displacement estimate feedback yields better results than the acceleration feedback at lower disturbance frequencies, both perform relatively similarly at higher frequencies.
% %
% Since the displacement estimate filter was tuned accounting for lower disturbance frequencies, an improved performance may be obtained by tuning the filter for higher frequencies.
Results show that this pseudo-displacement feedback outperforms pure acceleration feedback at lower disturbance frequencies, whereas both approaches perform comparably at higher frequencies.
Since the displacement-estimate filter was tuned primarily for lower-frequency disturbances, further performance gains are expected by optimizing the filter design for higher-frequency operation.
Future work will focus on enhancing the pseudo-displacement feedback scheme and reducing the dependence of RCAC performance on hyperparameter tuning.
% Future work will focus on improving the performance of the pseudo-displacement estimate feedback, as well as reducing the technique's dependence on $d_\rmf$ and $R_u$ hyperparameter tuning.

\bibliography{bib_paper}
\end{document}

%% file: PackagesCommands.tex
\usepackage{amsmath} % assumes amsmath package installed
\usepackage{amsfonts}
\usepackage{float} 
\usepackage{subcaption}

\usepackage{tikz}
\usetikzlibrary{circuits.ee.IEC}
\usetikzlibrary{shapes,arrows,calc,positioning}
\tikzstyle{bigblock} = [draw, fill=blue!20, rectangle, 
    minimum height=6em, minimum width=8em]
\tikzstyle{medblock} = [draw, fill=blue!20, rectangle, 
    minimum height=4em, minimum width=4em]    
\tikzstyle{mux} = [draw, fill=black!20, rectangle, 
    minimum height=5em, minimum width=0.1em]    
\tikzstyle{smallblock} = [draw, fill=blue!20, rectangle, 
    minimum height=2em, minimum width=3em]
    
\tikzstyle{data_block} = [draw, fill=green!20, rectangle, 
    minimum height=2em, minimum width=3em]
\tikzstyle{ops_block} = [draw, fill=blue!20, rectangle, 
    minimum height=2em, minimum width=3em]    
\tikzstyle{est_block} = [draw, fill=red!20, rectangle, 
    minimum height=2em, minimum width=3em]    
    
\tikzstyle{sum} = [draw, fill=blue!20, circle, node distance=1cm,minimum height=0.5cm]
\tikzstyle{signal} = [coordinate]
\tikzstyle{pinstyle} = [pin edge={to-,thin,black}]
\tikzstyle{block} = [draw, fill=blue!20, rectangle, 
    minimum height=3em, minimum width=9em]
\tikzstyle{blockS} = [draw, fill=blue!20, rectangle, 
    minimum height=3em, minimum width=4em]    
\tikzstyle{input} = [coordinate]
\tikzstyle{output} = [coordinate]

\tikzstyle{gain} = [draw, fill=blue!20, regular polygon, regular polygon sides=3, shape border rotate=30]
\tikzstyle{gain_vert} = [draw, fill=blue!20, regular polygon, regular polygon sides=3, shape border rotate=0]

\usetikzlibrary{matrix}

\usetikzlibrary{positioning}
\usetikzlibrary{math}

\usepackage{hyperref}
\usepackage{xcolor}
\hypersetup{
    colorlinks,
    linkcolor={blue!100!black},
    citecolor={blue!50!black},
    urlcolor={blue!80!black}
}

\newcommand{\bc}{\begin{center}}
\newcommand{\ec}{\end{center}}
\newcommand{\benum}{\begin{enumerate}}
\newcommand{\eenum}{\end{enumerate}}

\newcommand{\matl}{\left[ \begin{array}}
\newcommand{\matr}{\end{array} \right]}

\renewcommand{\matl}{\begin{bmatrix}}
\renewcommand{\matr}{\end{bmatrix}}

\newcommand{\matls}{\left[ \begin{smallmatrix}}
\newcommand{\matrs}{\end{smallmatrix} \right]}
\newcommand{\isdef}{\stackrel{\triangle}{=}}

\newcommand{\rmR}{{\rm R}}

\newcommand{\rmT}{{\rm T}}

\newcommand{\rmb}{{\rm b}}
\newcommand{\rmc}{{\rm c}}
\newcommand{\rmd}{{\rm d}}

\newcommand{\rmf}{{\rm f}}
\newcommand{\rmg}{{\rm g}}

\newcommand{\rmm}{{\rm m}}

\newcommand{\rms}{{\rm s}}

\newcommand{\rmu}{{\rm u}}

\newcommand{\rmy}{{\rm y}}

\newcommand{\BBC}{{\mathbb C}}

\newcommand{\BBR}{{\mathbb R}}

\newcommand{\BBZ}{{\mathbb Z}}

\newcommand{\bfq}{{\bf q}}

\newcommand{\SC}{{\mathcal C}}

\newcommand{\SF}{{\mathcal F}}

\newcommand{\SM}{{\mathcal M}}

\usepackage{enumitem,amssymb}
\newlist{todolist}{itemize}{2}
\setlist[todolist]{label=$\square$}
\usepackage{pifont}

\DeclareMathOperator{\vek}{vec}